\documentclass[11pt, aas_macros]{article} 
\pdfoutput=1

\usepackage[utf8]{inputenc} 
\usepackage{jcappub}

\usepackage{rotating} 
\usepackage{multirow}
\usepackage{graphicx} 
\usepackage{subfig}

\usepackage{booktabs} 
\usepackage{array} 
\usepackage{paralist} 
\usepackage{verbatim} 
\usepackage{subfig} 
\usepackage{epsfig}
\usepackage{amsmath}

\usepackage{fancyhdr} 
\usepackage{sectsty}
\allsectionsfont{\sffamily\mdseries\upshape} 

\usepackage[nottoc,notlof,notlot]{tocbibind} 
\usepackage[titles,subfigure]{tocloft} 

\newlength{\figlength}
\usepackage{soul}

\usepackage{hyperref}

\newcommand{\adsurl}[1]{\href{#1}{ADS}} 
\providecommand{\url}[1]{\href{#1}{#1}}

\title{A Search for Correlation of Ultra-High Energy Cosmic Rays with IRAS-PSCz and 2MASS-6dF Galaxies}
\author[a, b]{Foteini Oikonomou,}
\author[c]{Amy Connolly,}
\author[a]{Filipe B. Abdalla}
\author[a]{Ofer Lahav,}
\author[a, b]{Shaun A. Thomas,}
\author[d]{David Waters}
\author[e]{and Eli Waxman}

\affiliation[a]{Astrophysics Group, Department of Physics and Astronomy, University College London,\\
Gower Street, London WC1E 6BT, United Kingdom}
\affiliation[b]{UCL - Institute of Origins, Faculty of Mathematical and Physical Sciences, University College London}
\affiliation[c]{Department of Physics and CCAPP, The Ohio State University,\\
191 West Woodruff Avenue, Colombus, Ohio 43210, USA}
\affiliation[d]{High Energy Physics Group, Department of Physics and Astronomy, University College London}
\affiliation[e]{Department of Particle Physics \& Astrophysics, Weizmann Institute of Science, \\
Rehovot 76100, Israel}

\emailAdd{fotini@star.ucl.ac.uk}

\abstract{
\noindent We study the arrival directions of 69 ultra-high energy cosmic rays (UHECRs) observed at the Pierre Auger Observatory (PAO) with energies exceeding $55$~EeV. We investigate whether the UHECRs exhibit the anisotropy signal expected if the primary particles are protons that originate in galaxies in the local universe, or in sources correlated with these galaxies. We cross-correlate the UHECR arrival directions with the positions of IRAS-PSCz and 2MASS-6dF galaxies taking into account particle energy losses during propagation. This is the first time that the 6dF survey is used in a search for the sources of UHECRs and the first time that the PSCz survey is used with the full 69 PAO events. The observed cross-correlation signal is larger for the PAO UHECRs than for 94\% (98\%) of realisations from an isotropic distribution when cross-correlated with the PSCz (6dF). On the other hand the observed cross-correlation signal is lower than that expected from $\gtrsim 85\%$ of realisations, had the UHECRs originated in galaxies in either survey.  The observed cross-correlation signal does exceed that expected by $50\%$ of the realisations if the UHECRs are randomly deflected by intervening magnetic fields by $5^{\circ}$ or more. We propose a new method of analysing the expected anisotropy signal, by dividing the predicted UHECR source distribution into equal predicted flux radial shells, which can help localise and constrain the properties of UHECR sources. We find that the 69 PAO events are consistent with isotropy in the nearest of three shells we define, whereas there is weak evidence for correlation with the predicted source distribution in the two more distant shells in which the galaxy distribution is less anisotropic.
}


\arxivnumber{1207.4043}

\date{\today} 

\begin{document}
\maketitle
\flushbottom
\section{Introduction}
\label{sec:Introduction}

The origin of cosmic rays (CRs) with energies exceeding $10$~EeV ($1$~EeV $= 10^{18}$~eV) remains unknown despite decades of extensive research. Such particles are referred to as ultra-high energy cosmic rays (UHECRs) and their sources are probably extra-galactic \cite{2000PhR...327..109B}, \cite{2000RvMP...72..689N}. Their gyro-radius is larger than could be contained in the magnetic field of our galaxy \cite{1984ARA&A..22..425H}.\\
\indent UHECRs with energy above $\sim 50$~EeV are above the threshold for pion photo-production upon collision with cosmic microwave background (CMB) photons, a process which was predicted soon after the discovery of CMB radiation, known as the GZK cutoff \cite{1966PhRvL..16..748G}, \cite{Zatsepin-Kuzmin}. The most recent published cosmic ray spectra measured by the High Resolution Fly's Eye (HiRes), PAO and Telescope Array (TA) experiments confirm this cutoff in the CR spectrum \cite{2008PhRvL.100j1101A}, \cite{2010PhLB..685..239A}, \cite{2012arXiv1205.5067A}. \\
\indent As a result of the GZK process UHECRs that arrive on earth with energy exceeding $50$~EeV must originate in sources within a few hundred mega-parsecs (Mpc). If UHECRs are protons and extra-galactic magnetic fields (EGMFs) are not too large, observed UHECRs must point back to their sources within a few degrees. Further, if UHECRs originate in some astrophysical population, their arrival direction distribution should be correlated with that population as well as with the distribution of large scale structure (LSS) in the local universe, since matter in the Universe is clustered \cite{1997ApJ...483....1W}. \\
\indent Heavier UHECR nuclei with energy around $50$~EeV would have their arrival directions smeared by intervening magnetic fields. Other models of particle physics beyond the Standard Model have also been proposed for the origin of UHECRs (so called $top-down$ models, see \cite{2000PhR...327..109B} for a review).\\
\indent Only a handful of known astrophysical populations are likely to have the required power to accelerate CRs to such high energies \cite{1984ARA&A..22..425H}, \cite{2010PhyU...53..691P}. The most likely candidates are active galactic nuclei (AGN) and gamma ray bursts (GRBs) (see however \cite{2012Natur.484..351A}, \cite{2012arXiv1205.3479D}, \cite{2012ApJ...752...29H} for a recent result and alternative interpretations on GRBs as UHECR sources). \\
\indent Besides the origin, another question which remains open is the composition of UHECRs. The PAO, the largest UHECR observatory to date, has not reported on the composition of UHECRs with energy exceeding $20$~EeV, presumably due to the small number of observed events. Their most recent results between $1- 20$~EeV suggest that there is a smooth transition from light, proton-like to heavier, Fe-like nuclei \cite{2010PhRvL.104i1101A}, whereas HiRes and TA results up to an energy of $50$~EeV agree with a proton-like composition \cite{2010arXiv1010.2690S}, \cite{Jui:2011vm}. There are a number of reasons why one wouldn't expect heavy nuclei at energies above the GZK threshold (see for example \cite{2011arXiv1101.1155W}) not least that the leading extra-galactic candidate sources are expected to accelerate primarily protons. \\
\indent In this work we revisit the question of the origin of UHECRs, following the release of the arrival directions of 69 UHECRs with energy above $55$~EeV recorded until December 2009 by the PAO \cite{2010APh....34..314T}. We model the local UHECR source distribution using galaxy catalogues of the nearby universe, namely the six degree Field Galaxy Survey (6dF), which is being used in this work for the first time to derive the expected UHECR source distribution, and the IRAS Point Source Catalogue of redshifts (PSCz), which is being used here for the first time to analyse the updated dataset of 69 UHECRs. We cross-correlate the arrival directions of the observed UHECRs with the predicted source distribution to assess whether a correlation exists, using the statistic $X$ proposed in \cite{kashti2008}. We propose a new method of studying any correlation between UHECR arrival directions and model source distribution, by dividing the predicted UHECR source distribution into radial shells with distance, which can help localise the sources of UHECRs if the expected correlation between the sources and arrival directions exits. Finally,  we investigate the effect of random magnetic deflections on any correlation between the source population and the observed UHECR arrival direction distribution. Throughout this work we assume a flat universe with $\Omega_{M}=0.25$, $\Omega_{\Lambda}=0.75$ and $H_o$ = 70 km s$^{-1}$ Mpc$^{-1}$.\\
\indent A number of authors have looked for the sources of UHECRs in the past, by cross-correlating the observed arrival directions with the positions of nearby galaxy catalogues, in particular the PSCz \cite{kashti2008}, \cite{2010ApJ...716..914B}, \cite{2009JCAP...06..031T}, \cite{2006JCAP...01..009C}, \cite{2009JCAP...04..003K}, 2MRS \cite{2010ApJ...713L..64A},  \cite{2010APh....34..314T}, \cite{2009arXiv0906.2347T} and others with the positions of specific classes of objects, AGN \cite{2007Sci...318..938T}, \cite{2008APh....29..188P}, \cite{george_et_al}, \cite{2009PhRvD..80l3018P}, \cite{Mortlock}, BL Lacertae objects (BL Lacs)  \cite{2001JETPL..74..445T} and luminous infrared galaxies (LIRGs) \cite{2010ApJ...716..914B}, reporting different degrees of correlation depending on the UHECR sample, statistical approach and source population used. The authors of \cite{kashti2008} analysed the arrival directions of 27 UHECRs with energy above $55$~EeV detected until August 2007 and found that the UHECRs exhibited a stronger correlation with propagation weighted PSCz galaxies than $99.8\%$ of isotropic realisations. The authors of \cite{2010APh....34..314T} analysed the distribution of the 69 observed UHECRs and found that 21 of the 55 that survive their cuts correlate with nearby AGN. The probability of finding such a correlation assuming isotropy is 0.003. Further they cross-correlated the arrival directions of the UHECRs with the positions of 2MRS galaxies and Swift-BAT X-ray sources \cite{2010ApJS..186..378T}. For the values of the free parameters in their models that maximise the likelihood they found that the fraction of isotropic realisations that yield a higher likelihood than the observed UHECRs are 0.004 and $2 \times 10^{-4}$ for the 2MRS and Swift-BAT respectively. The results they obtained are \textit{a posteriori} and do not constitute a confidence level on anisotropy. Because of the different approaches followed their results are not directly comparable to ours. \\
\indent This work is organised as follows. In Section \ref{sec:Methodology} we present the formalism used to analyse the arrival direction distribution of UHECRs, in Section \ref{sec:Results} we present our results and in Section \ref{sec:Discussion} we discuss the implications of our results and conclude.

\section{Methodology}
\label{sec:Methodology}
 \begin{figure}
\centering
\includegraphics[width=1.55\figlength, height=0.4\textheight]{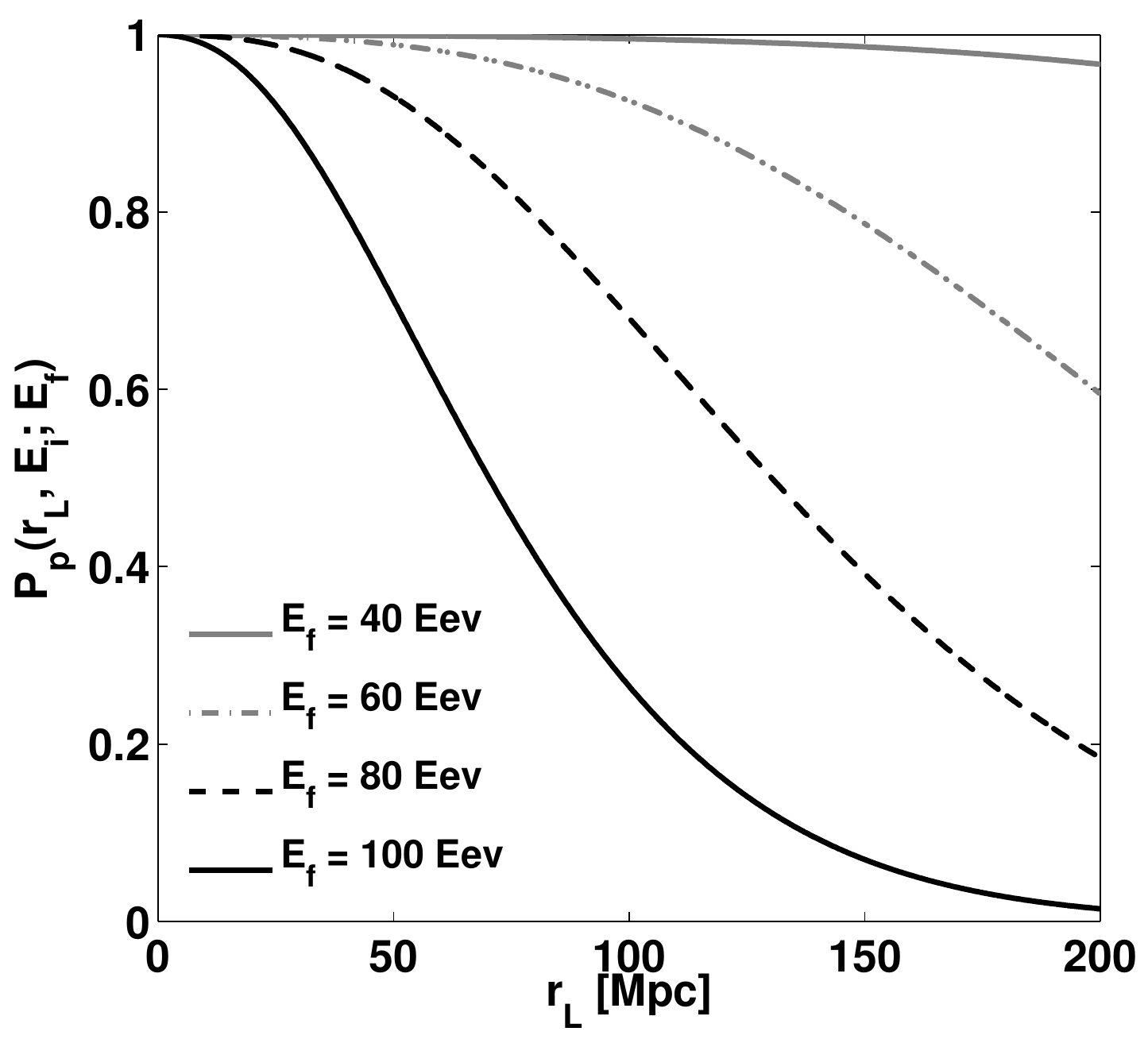}
\caption{The function $P_p(r_L, E_i; E_f)$, which represents the probability that a UHECR emitted by a source at luminosity distance $r_L$, with initial energy $E_i$, will be observed with final energy above $E_f$, here shown as a function of propagation distance for  $E_i = 200$~EeV and $E_f \geq 40, 60, 80, 100$~EeV.}
\label{fig:fodor_diagnostic}
\end{figure}
\indent We study the correlation between the arrival directions of the 69 PAO observed events ($\S$\ref{subsec:auger}) and the distribution of matter in the local LSS which we model using the 6dF ($\S$\ref{subsec:6df}) and the PSCz ($\S$\ref{subsec:pscz}) catalogues. In $\S$\ref{subsec:gzk_weight} we present our models of UHECR source distribution. Throughout this analysis we take into account the non-uniform PAO exposure ($\S$\ref{subsec:auger}). In $\S$\ref{subsec:magnetic_fields} we explain our treatment of the expected UHECR random magnetic deflections. In $\S$\ref{subsec:statistic} we present our statistical approach.
\subsection{The Pierre Auger observatory}
\label{subsec:auger}
\indent The PAO, in Malargue Argentina is a cosmic ray observatory dedicated to the detection of cosmic rays of energy greater than of $1$~EeV. Data taking started in January 2004 and since then the arrival direction and reconstructed energy of 69 UHECRs exceeding $55$~EeV have been published. At the PAO, which has a detection area of $3000$~km$^2$, UHECR particle showers are detected by 1600 ground based Cherenkov detectors surrounded by 24 fluorescence telescopes, which measure the amount of energy dissipated in the atmosphere in the form of ultra-violet radiation \cite{2004NIMPA.523...50A}. The combination of these two techniques provides the most accurate reconstruction of UHECR shower geometry to date. Full time operation ensures an exposure which is uniform in right ascension and fully efficient for zenith distance up to $\theta_{m} = 60^{\circ{}}$. The PAO exposure is a function of declination and is given by \cite{sommers2001}:
\begin{equation}
\label{eq:PAO_exposure}
\omega(\delta) \propto \cos(a_0) \cos(\delta) \sin(\alpha_m) + \alpha_{m} \sin(a_0) \sin(\delta),
\end{equation}
\noindent where $\delta$ is the declination and $a_0$ is the latitude of the PAO which is $-35.2^{\circ{}}$. Here, $\alpha_m$ is given by: $$\alpha_m = \begin{cases}0 & \mbox{if $\xi > 1$} \\ \pi  & \mbox{if $\xi < -1$}  \\ \cos^{-1}(\xi) & \mbox{otherwise} \end{cases}$$
and $$\xi \equiv \frac{\cos(\theta_m) - \sin(a_0)\sin(\delta)}{\cos(a_0) \cos(\delta)}.$$ The PAO has an integrated exposure of $\sim 20000$~$\textrm{km}^2$~$\textrm{year}$ and $\sim 20$ UHECRs above the GZK threshold are detected every year since its completion in November 2008 \cite{2011arXiv1107.4809T}. 
\subsection{The 6dF}
\label{subsec:6df}
\indent The 6dF is a redshift and peculiar velocity survey of 2MASS selected galaxies which was carried out using the Six-Degree Field instrument on the Schmidt Telescope of the Anglo-Australian Observatory \cite{the6df}. The survey, which covers the entire Southern Sky (excluding the Galactic plane, $|b| > 10^{\circ{}}$), resulted in a catalogue of 125,071 galaxies and 110,256 associated redshifts. The 6dF has median redshift $\overline{z}=0.053$ which corresponds to a comoving distance of 225 Mpc in the cosmological model we are assuming. The 6dF field of view covers 80\% of the PAO field of view by area. Taking into account the total declination dependent PAO acceptance \eqref{eq:PAO_exposure}, which is smaller for positive declinations, it covers 86.2\% of the instantaneous PAO exposure. The 6dF is near-infrared selected, which means it favours older, bulge-dominated galaxies and therefore it is minimally affected by dust extinction. In this work the final 6dF data release \cite{the6df} is being used for the first time for a study of the origin of high energy particles. 
\subsection{The IRAS PSCz}
\label{subsec:pscz}
\indent The PSCz is a redshift catalogue of infrared selected IRAS galaxies which covers 84\% of the sky \cite{pscz}. It contains 14,677 galaxies with associated redshifts, and median redshift  $(\overline{z}=0.028)$ which corresponds to a comoving distance of 119 Mpc. Since the galaxies in the PSCz are infrared selected there is a preference for young, star-forming galaxies in the catalogue.\\
\indent We have chosen to use two complementary galaxy surveys to derive the expected UHECR source distribution. The PSCz is a shallow nearly full-sky galaxy survey and its use facilitates comparison with results of previous studies. The 6dF on the other hand is a much larger survey ($\sim 20$~times more galaxies than the PSCz in the southern hemisphere). The different median depths of the 2 surveys mean that they highlight different structures of the nearby universe (for example the Shapley Concentration (centred at $(l \sim -50^{\circ}, b \sim 30^{\circ})$ at a distance $\sim 200$~Mpc is prominent in the 6dF) although there is a significant overlap. The 2 surveys consist of different galaxy populations with slightly different clustering properties. This means that they can be used to distinguish between different astrophysical populations as UHECR sources, although this is a subtle difference that will require a large UHECR dataset and a better understanding of UHECR magnetic deflections before it can be pursued. 
\subsection{Model of UHECR source distribution}
\label{subsec:gzk_weight}
We use the galaxy catalogues introduced above to model a UHECR source population which is steady and follows the distribution of luminous matter in the local universe. We consider a model in which the local number density of UHECR sources is comparable to that of galaxies in the local Universe ($\overline{n}_0 = 10^{-2}$~$\textrm{Mpc}^{-3}$) and in which individual UHECR sources are faint i.e. each source produces 1 or no events and the probability of a single source producing multiple events is low. Further we assume that all UHECR sources are intrinsically identical.\\
\indent It is also possible that the observed UHECR flux is dominated by a few ``bright'' sources (such as in models in which the nearby radio galaxy Centaurus A is responsible for much of the observed UHECR flux \cite{1978A&A....65..415C}, \cite{1996APh.....5..279R}, \cite{2001PhRvL..87h1101A}).  The number of ``repeaters'', i.e. groups of UHECRs with arrival directions separated by less than a few degrees that may be associated with a single source, can constrain the density of sources of UHECRs \cite{1997ApJ...483....1W}, \cite{2000PhRvL..85.1154D}, \cite{2001PhRvD..63b3002F}. For the sole purpose of estimating the source density, we have searched for repeaters separated by less than $3^{\circ}$, intended to reflect the uncertainty associated with instrumental resolution ($\sim 1^{\circ}$) and possible magnetic deflections of a few degrees (see $\S$\ref{subsec:magnetic_fields}), among the 69 observed PAO events. We found 4 pairs of events, and using the procedure outlined in \cite{2000PhRvL..85.1154D} we estimate that this implies $\sim 1800$ UHECR sources. Using a horizon of 100 Mpc for 60 EeV UHECRs, we find that the absolute minimum source density is $\frac{1800~\rm{ sources}}{\frac{4 \pi}{3} \cdot (100~\rm{Mpc})^3} \sim 4 \times 10^{-4}$~ Mpc$^{-3}$. If the observed number of pairs arrises by chance, which is expected to happen $22\%$ of the time for an isotropic source distribution, the source density could be much larger than this. \\
\indent We assign each galaxy in the model source distribution a weight, proportional to the expected UHECR flux from that source, which depends on the flux suppression with distance and the energy with which protons are emitted (this initial energy determines the energy losses during propagation due to interaction with background photons). The flux weight for a galaxy at luminosity distance $r_L$ from earth that emits a cosmic ray proton with initial energy $E_i$ which reaches the earth with energy equal to or greater than $E_f$, is given by:
\begin{equation}
\label{eq:weight}
\omega(r_L)_{\rm{flux}} = \frac{1}{{r_L}^2} \int^{E_{\rm{i, max}}} _{E_{f}'} dE_i \int^{E_{i}} _{E_{f}'} dE_f  \rho_p(r_L, E_i; E_f) I(E_i),
\end{equation}
\noindent where
\begin{equation}
\rho_p(r_L, E_i; E_f) = \left| \frac{\partial P_p(r_L, E_i; E_f)}{\partial E_f} \right|,
\end{equation}
\noindent is the derivative of the function $P_p(r_L, E_i; E_f)$ which gives the probability of a proton arriving on earth with energy $E_f$ if it was emitted with energy $E_i$ by a source at distance $r_{L}$, introduced in \cite{2000ApJ...542..542B}. For $E_{f}'$, the final energy of the 69 PAO UHECRs, we conservatively adopt the lowest measured energy (55 EeV) present in the PAO dataset. We set $E_{\rm{i, max}}$ which is the maximum energy achievable through astrophysical processes equal to $10^{21}$~eV. The value of $E_{\rm{i, max}}$ is not fully constrained by observations but considering higher values of $E_{\rm{i, max}}$ would have a negligible effect on our results, since UHECRs above this energy promptly interact with background photons. The intrinsic spectrum of the UHECR sources is not yet known. The ultra-high energy part of the observed CR spectrum is well fit by a number of models (see for example \cite{2010JCAP...10..013K}). Here, we consider a power law spectrum
\begin{equation}
\label{eq:injection_spectrum}
 I(E_i)=I_0 E_i^\alpha \Theta(E_{\rm{i, max}}-E_i)
\end{equation}
\noindent with index $\alpha = -2$ which is what is expected in diffusive shock acceleration \cite{1987PhR...154....1B}, \cite{2001MNRAS.328..393A} and has been shown to agree with the observed PAO spectrum in \cite{2009JCAP...03..020K}. The step function restricts the UHECRs to have energy below $E_{\rm{i, max}}$. Here $I_0$ normalises the injected UHECR energy spectrum such that:
\begin{equation}
 \int^{E_{\rm{i, max}}}_{E_{f}} dE_i I_0 E_i^\alpha \Theta(E_{\rm{i, max}}-E_i) = 1.
\end{equation}
\noindent For the simulation of the proton propagation and energy losses we use the $P_p(r_L, E_i; E_f)$ function which was computed and made publicly available by the authors of \cite{2001PhRvD..63b3002F} and \cite{2003JCAP...11..015F} at \url{http://www.desy.de/~uhecr/P_proton}. In figure \ref{fig:fodor_diagnostic} we plot the function $P_p(r_L, E_i; E_f)$ for UHECRs with $E_i = 200$~EeV and $E_f \geq 40, 60, 80, 100$~EeV. \\
\indent To account for the effect of the flux limit of the survey on the observed number density of galaxies as a function of distance we need the survey's selection function, defined as the expected number density of galaxies in the survey as a function of distance, in the absence of clustering. One can model the selection function using a fit to the survey's redshift distribution, often parametrised as: 
\begin{equation}
dN(z) = A z^{\beta} \exp \left[ - \left( \frac{z}{z_p} \right) ^ {\gamma} \right] dz,
\end{equation}
\noindent where $A$ gives the normalisation, $z_p$ the peak of the distribution and $\beta$, $\gamma$ control the slope. All 4 parameters are specific to the survey (see \cite{the6df} and \cite{pscz} for 6dF and PSCz redshift distributions respectively). The overall selection function $\psi(r_{c})$ is the redshift distribution divided by the volume element \cite{2006MNRAS.373...45E}:
\begin{equation}
\label{eq:generic_sel_fn}
\psi(r_c) = \frac{1}{\Omega_s r^2_c} \left( \frac{dN}{dz}\right)_{r_{c}}\left(\frac{dz}{dr_c}\right)_{r_{c}},
\end{equation}
where $\Omega_s$ is the solid angle of the survey and $r_c$ the comoving distance. The function $\psi(r_{c})$ is normalised to the value it takes at some small distance $r_{\rm{min}}$ below which we believe the survey includes all existing galaxies. The value of $r_{\rm{min}}$ is not very well constrained by observations. We choose the minimum redshift found in each survey $\approx 1$~Mpc as the default $r_{\rm{min}}$, but in $\S$\ref{subsec:systematics} we study the sensitivity of our results to the choice of $r_{\rm{min}}$.\\
\indent We weight the observed galaxy distribution by the inverse of $\psi(r_{c})$, so that the effective contribution of each survey galaxy to the model source distribution is: 
\begin{equation}
\omega_{\rm{gal}} = \frac{\omega(r_L)_{flux}}{\psi(r_{c})}.
\label{eq:gal_weight}
\end{equation}
\indent Galaxy surveys measure redshifts not distances, and peculiar velocities along the line-of-sight affect our estimates of actual distances. The redshift of an object is defined by:
\begin{equation}
cz = H_{0} r + \mathbf{\left( v \left(r \right)- v\left( 0 \right) \right) \cdot \hat{r}},
\end{equation}
\noindent where $\mathbf{v \left(r \right)}$ is the object's peculiar velocity and  $\mathbf{v \left(0 \right)}$ is the observer's peculiar velocity. Working in a reference frame where $\mathbf{\Delta v = v \left(r \right)- v\left( 0 \right)}$ is small allows more accurate distance estimates. In the local universe, out to $cz \sim 3000$~kms$^{-1}$, where galaxies share the motions of the Local Group (LG) it is best to convert to the LG rest frame. Further away galaxy peculiar velocities are independent of the LG velocity and $\mathbf{\Delta v}$ is smaller in the CMB rest frame. In this work we are interested in the distribution of matter nearby, hence it is most useful to work with LG frame redshifts. \\
\indent To convert between $cz_{\rm{Hel}}$, that is redshift measured in the heliocentric frame, and $cz_{LG}$ we use the relation \cite{1999AJ....118..337C}:
\begin{equation}
cz_{LG} = cz_{\rm{Hel}} - 79 \cos(l) \cos(b)+296 \sin(l) \cos(b)-36 \sin(b),
\end{equation}
\noindent where $l$ and $b$ are the Galactic longitude and latitude respectively. 
\subsection{Magnetic fields}
\label{subsec:magnetic_fields}
A relativistic particle of energy $E$ and charge $Z$ in a magnetic field has a Larmor radius given by $r_{Lar}=\frac{E}{ZeB_{\perp}}$, where $B_{\perp}$ is the field strength in the direction perpendicular to the momentum of the particle. A relativistic proton propagating a distance $D$ in a magnetic field with correlation length $\lambda$ will suffer a deflection \cite{1996ApJ...472L..89W}:
\begin{equation}
\theta(E,D) \approx 0.5^{\circ} \left(\frac{D}{\lambda} \right)^{\frac{1}{2}} \left( \frac{E}{100~\textrm{EeV}}\right)^{-1} \left( \frac{\lambda}{1~\textrm{Mpc}}\right)\left( \frac{B}{10^{-9}~\textrm{G}}\right).
\end{equation} 
The strength and distribution of EGMFs are poorly known (see reviews of existing data in \cite{1994RPPh...57..325K}, \cite{1997FCPh...19....1V}). Dense large scale structures such as galaxy clusters and galaxy filaments are likely to support relatively large magnetic fields whereas in between structures EGMFs are probably negligible. Large scale simulations of EGMFs \cite{2005JCAP...01..009D}, as well as results from a recent semi-analytic analysis \cite{2008PhRvD..77l3003K}, conclude that deflections of proton UHECRs of energy E $\geq 40$~EeV do not exceed $3^{\circ}$ over $99\%$ of the sky for  propagation distance $\sim 100$~Mpc.\\
\indent Magnetic deflections suffered by heavier UHECR nuclei are expected to be much larger and can completely wash out the directional correlation of UHECRs with their sources \cite{2009arXiv0907.5194A}. The composition of the UHECR sample we are considering here is at present uncertain but the anisotropies we would expect at lower energies if this dataset was composed of nuclei rather than nucleons have not been observed (see \cite{2009JCAP...11..009L}, \cite{2011arXiv1106.3048T}). Throughout most of this analysis we consider proton UHECRs. In order to bracket possible proton UHECR magnetic deflections we perform our analysis by averaging over angular bins in the range $3.9^{\circ} - 7.3^{\circ}$ (see details of our binning method in $\S$\ref{subsec:statistic}). We exclude the Galactic plane from our analysis, $|b| \geq 12^{\circ}$, since UHECRs travelling through the Galactic plane may suffer strong deflections due to the magnetic field in the Galactic disc. In $\S$\ref{subsec:deflection} we relax the assumption of nucleonic composition and investigate the effect that larger random deflections have on any correlation between the source population and the observed arrival direction distribution. 
\subsection {Correlation statistic} 
\label{subsec:statistic}
To detect any existing anisotropy signal we divide the sky into equal area bins and consider the counts-in-cells statistic $X$, proposed in \cite{kashti2008}, which characterises the correlation between the predicted and observed UHECR arrival direction distribution. In \cite{kashti2008} $X$ was shown to be more sensitive to the expected anisotropy signal than the angular power spectrum (e.g. \cite{sommers2001}, \cite{1996ApJ...468..214T}) and the two point correlation function \cite{2006APh....26...10K}, which are other statistical measures commonly used in clustering analyses. Since the exact magnetic delfection suffered by proton UHECRs is unknown we consider bin-sizes in the range $3.9^{\circ} - 7.3^{\circ}$ to bracket expected UHECR deflections of a few degrees.\\
\indent Let $\omega_{\rm{Auger, i}}$ be the exposure of the PAO in bin $i$ \eqref{eq:PAO_exposure} and $\omega_{\rm{survey, i}}$ the weight imposed by the survey mask (given by ~\cite{the6df}, \cite{pscz}): $$\omega_{\rm{6dF, i}} = \begin{cases} 1 & \mbox{ for $|b| > 12^{\circ{}}$ and $\delta \leq 0^{\circ{}}$} \\ 0  & \mbox{otherwise} \end{cases}$$ $$\omega_{\rm{PSCz, i}} =  \begin{cases} 1 & \mbox{ for $|b| > 12^{\circ{}}$ } \\ 0  & \mbox{otherwise}. \end{cases}$$
\noindent Here $\left\{ i \right\}$ is the set of angular bins in the mask defined region. We assign each angular bin the value of $\omega_{\rm{Auger}}(\delta)$ at the centre of that bin and treat $\omega_{\rm{Auger}}$ as a constant within each bin. For the survey mask $\omega_{\rm{survey}}$, every bin that overlaps with the region excluded by the survey mask (even if it partly overlaps) is excluded.  \\
\indent To account for the combined effects of the survey mask and PAO exposure in bin $i$, we define $\omega_{\rm{exposure, i}}$ :
\begin{equation}
\omega_{\rm{exposure, i}} = \omega_{\rm{Auger, i}} \cdot \omega_{\rm{survey, i}}.
\end{equation}
We require that each of the quantities in our analysis are (a) weighted exactly once by the combined weight and (b) normalised so that each has a sum equal to the weighted sum of observed UHECRs in the mask defined region.\\ 
\indent We define $N_{\rm{CR, i}}$ to be the number of UHECRs detected in bin $i$. The number of observed PAO UHECRs in bin $i$ is related to $N_{CR,i}$ by:
\begin{equation}
N_{\rm{CR, i}} = \omega_{\rm{survey, i}} \cdot N_{\rm{Auger, i}},
\end{equation}
which effectively means that observed UHECRs outside the survey defined region are excluded. In the case where trial UHECR samples are drawn from the models of source distribution we have presented $N_{CR,i}$ is the number of UHECRs expected to be detected in bin $i$ in each model. \\
\indent The number of survey galaxies visible by the PAO is limited by the observatory's declination dependent exposure \eqref{eq:PAO_exposure}. To account for this effect, as well as to exclude those survey galaxies that lie near the Galactic plane where observations are not reliable, we subject each survey galaxy in bin $i$ to  the combined weight :
\begin{equation}
N_{\rm{gal, i}} = \omega_{\rm{exposure, i}} \cdot \sum_j \omega_{\rm{gal, j}},
\end{equation}
\noindent where the sum is over the weighted contribution to the flux of each survey galaxy $j$ \eqref{eq:gal_weight} in bin $i$. Note that the PAO exposure $\omega_{\rm{exposure, i}}$ is treated as a constant within a given angular bin and hence remains outside the sum. In the model where UHECR sources are correlated with galaxies in the nearby LSS, the expected number of cosmic rays $N_{\rm{M, i}}$ is simply $N_{\rm{gal, i}}$ normalised to the number of observed UHECRs according to condition (b):
\begin{equation}
N_{\rm{M, i}} = \frac{\sum_i N_{\rm{CR, i}}}{\sum_i N_{\rm{gal, i}}} \cdot N_{\rm{gal, i}}.
\end{equation}
\noindent In an isotropic model the expected number of cosmic rays is given by:
\begin{equation}
N_{\rm{iso, i}} = \frac{\omega_{\rm{exposure, i}}}{n} \cdot \left( \frac{4 \pi}{\int_0^\pi \! {2 \pi \sin(\delta) \omega_{\rm{Auger}}(\delta)  \ d\delta}} \cdot \sum_j N_{\rm{Auger, j}} \right),
\end{equation}
\noindent where $n$ is the number of angular bins in the sky. The integral $\int_0^\pi \! {2 \pi \sin(\delta) w(\delta)  \ d\delta}$ gives the total exposure of the PAO and hence  $\frac{4 \pi}{\int_0^\pi \! {2 \pi \sin(\delta) w(\delta)  \ d\delta}}$ ``unmasks'' $N_{\rm{Auger, i}}$, i.e. the quantity in the brackets gives the number of cosmic rays that would have been observed if the PAO had had a uniform, full sky acceptance.\\
\indent We generate mock realisations of UHECRs drawn from an isotropic distribution and cross-correlate them with the predicted source distribution based on the PSCz/6dF to model the distribution of values that the statistic $X$ takes in the case of an isotropic source distribution. Similarly we create mock realisations of UHECRs correlated with PSCz/6dF galaxies by sampling from a Poisson distribution with a mean equal to the number of UHECRs expected to be detected per unit area at the PAO above energy $E_{f}$ using the model of UHECR source distribution presented in $\S$\ref{subsec:gzk_weight}. Finally we compute the statistic for the observed PAO UHECRs, and determine which of the two models of source distribution is preferred by the observed PAO UHECR dataset. The statistic $X$ is defined as:
\begin{equation}
X = \sum_{i}\frac{\left( N_{\rm{CR, i}} - N_{\rm{iso, i}} \right) \cdot \left( N_{\rm{M, i}} - N_{\rm{iso, i}} \right)}{N_{\rm{iso, i}}}.
\label{eq:statistic}
\end{equation}

%
\section{Results}
\label{sec:Results}
\begin{figure}
\centering
\subfloat{\includegraphics[width = 1.25\figlength, height=0.18\textheight]{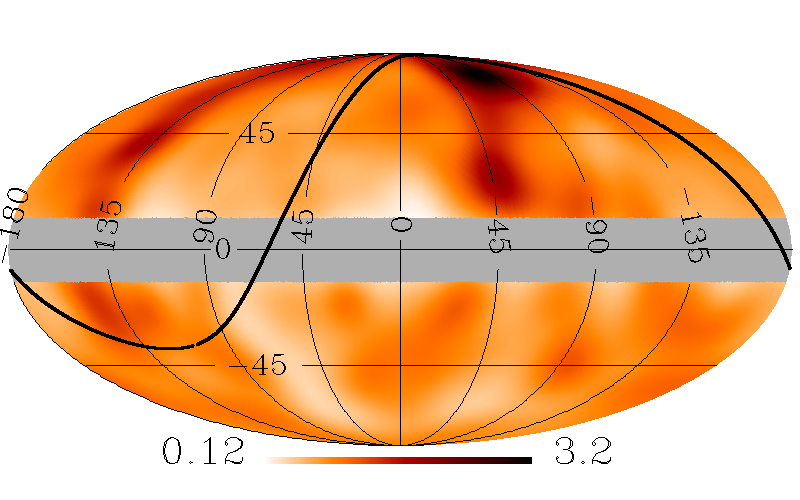}}
\subfloat{\includegraphics[width = 1.25\figlength, height=0.18\textheight]{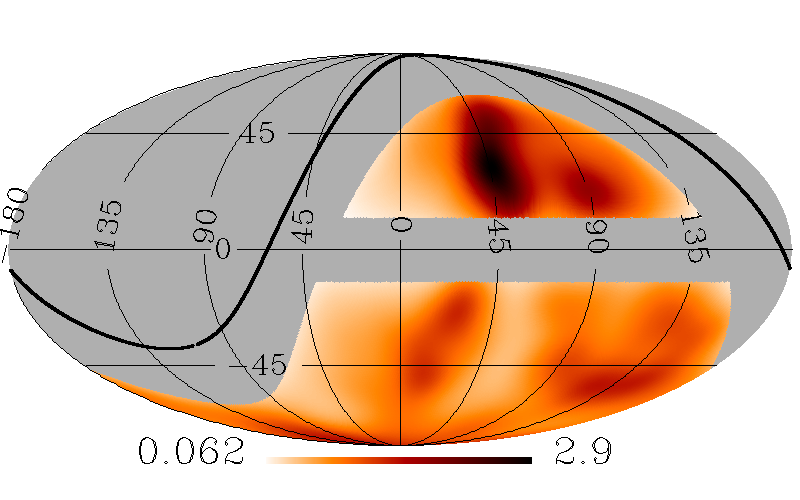}}\\
\caption{Maps of the predicted, local UHECR source distribution, in Galactic coordinates centred at the Galactic centre with the longitude $l$ increasing anti-clockwise, derived from the PSCz (left) and 6dF (right) for UHECRs with final energy $55$~EeV. The predicted source distribution has been smoothed with a Gaussian filter, with $\sigma = 7.2^{\circ}$ for presentation purposes. The intensity at each point is normalised to the average intensity in the map. The PAO is sensitive to the part of the sky below the thick black line. The regions in grey are excluded from our analysis (see details regarding galaxy survey masks $\S$\ref{subsec:statistic}).}
\label{fig:maps}
\subfloat{\includegraphics[width = 1.25\figlength, height=0.18\textheight]{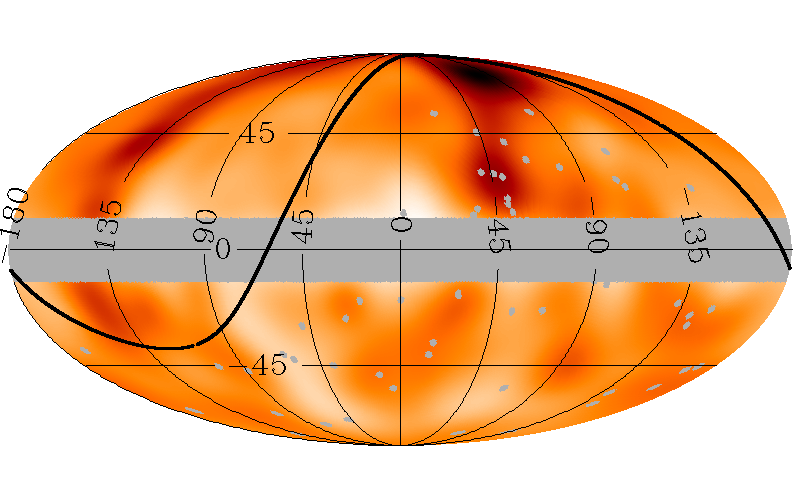}}
\subfloat{\includegraphics[width = 1.25\figlength, height=0.18\textheight]{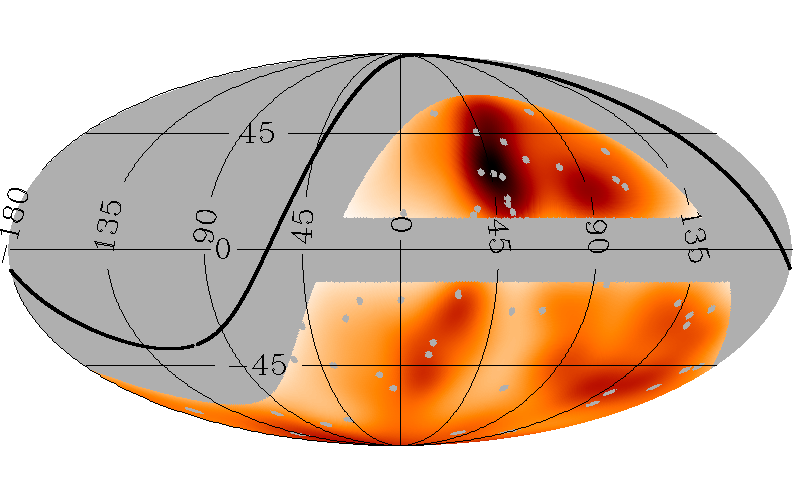}}\\
\caption{The predicted UHECR source distribution as in figure \ref{fig:maps} above, with those of the 69 PAO observed events whose arrival directions fall in the galaxy survey's field of view superimposed (in grey). The non-uniform PAO exposure has not been taken into account in these intensity maps, hence visual inspection of correlations can be misleading. A statistical analysis as in $\S$\ref{subsec:statistic} is required.}
\label{fig:maps_and_UHECRs}
\end{figure}
\begin{figure}
\centering
\includegraphics[width=2.5\figlength, height=0.8\textheight]{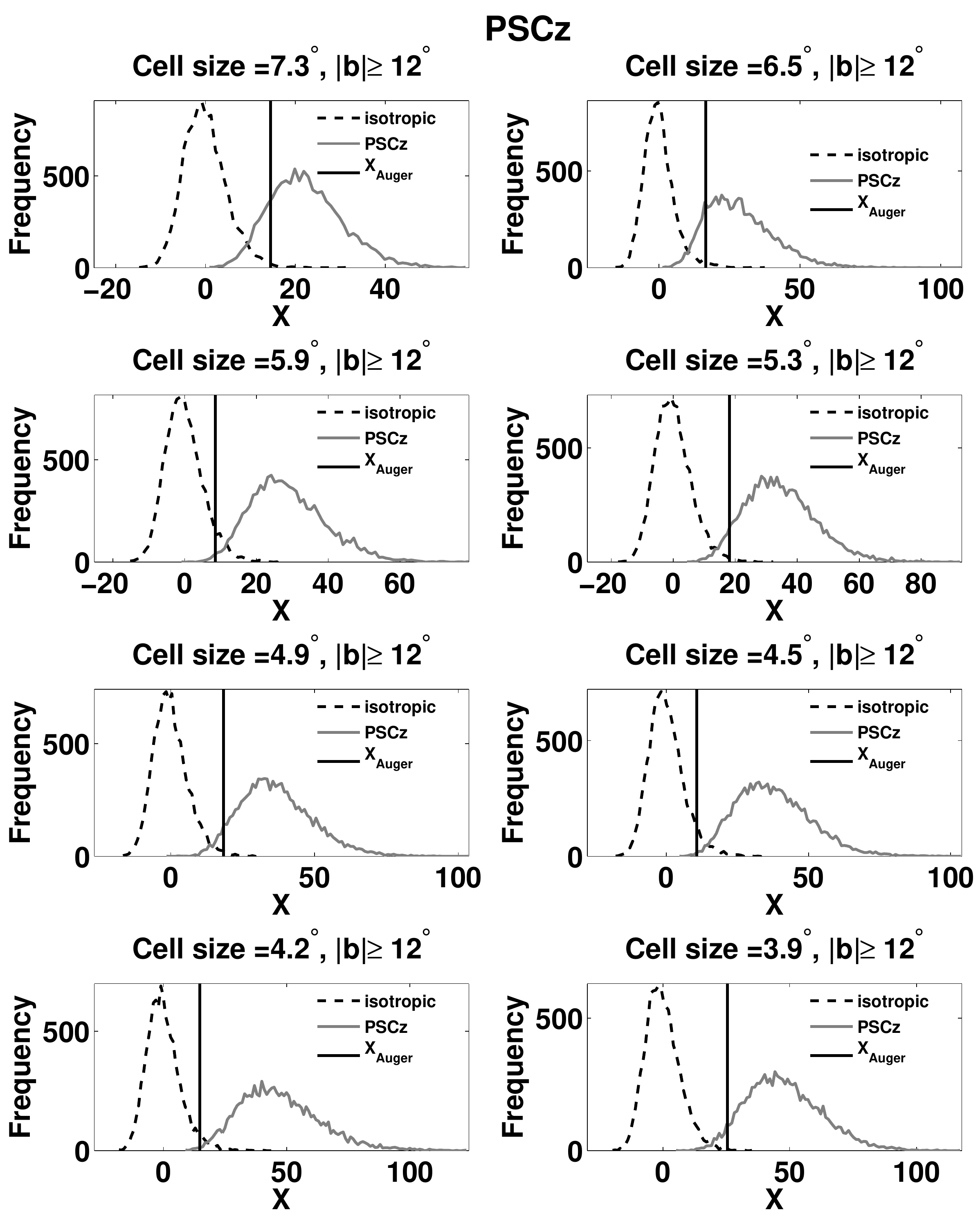}
\caption{The distribution of values of the correlation statistic $X$  \eqref{eq:statistic} in 10,000 mock realisations of a set of UHECRs drawn from an isotropic distribution (\textit{dashed histograms}) and from the model UHECR source distribution that follows the PSCz (\textit{solid histograms}). The value of $X$ obtained for the observed PAO events ($X_{\rm{Auger}}$) is given by the black solid line. Each subplot corresponds to a different cell size used for the counts-in-cells analysis in the range $7.3^{\circ}$ (top left) to $3.9^{\circ}$ (bottom right).}
\label{fig:histograms_cell_size_69_pscz}
\end{figure}
\begin{figure}
\centering
\includegraphics[width=2.5\figlength, height=0.8\textheight]{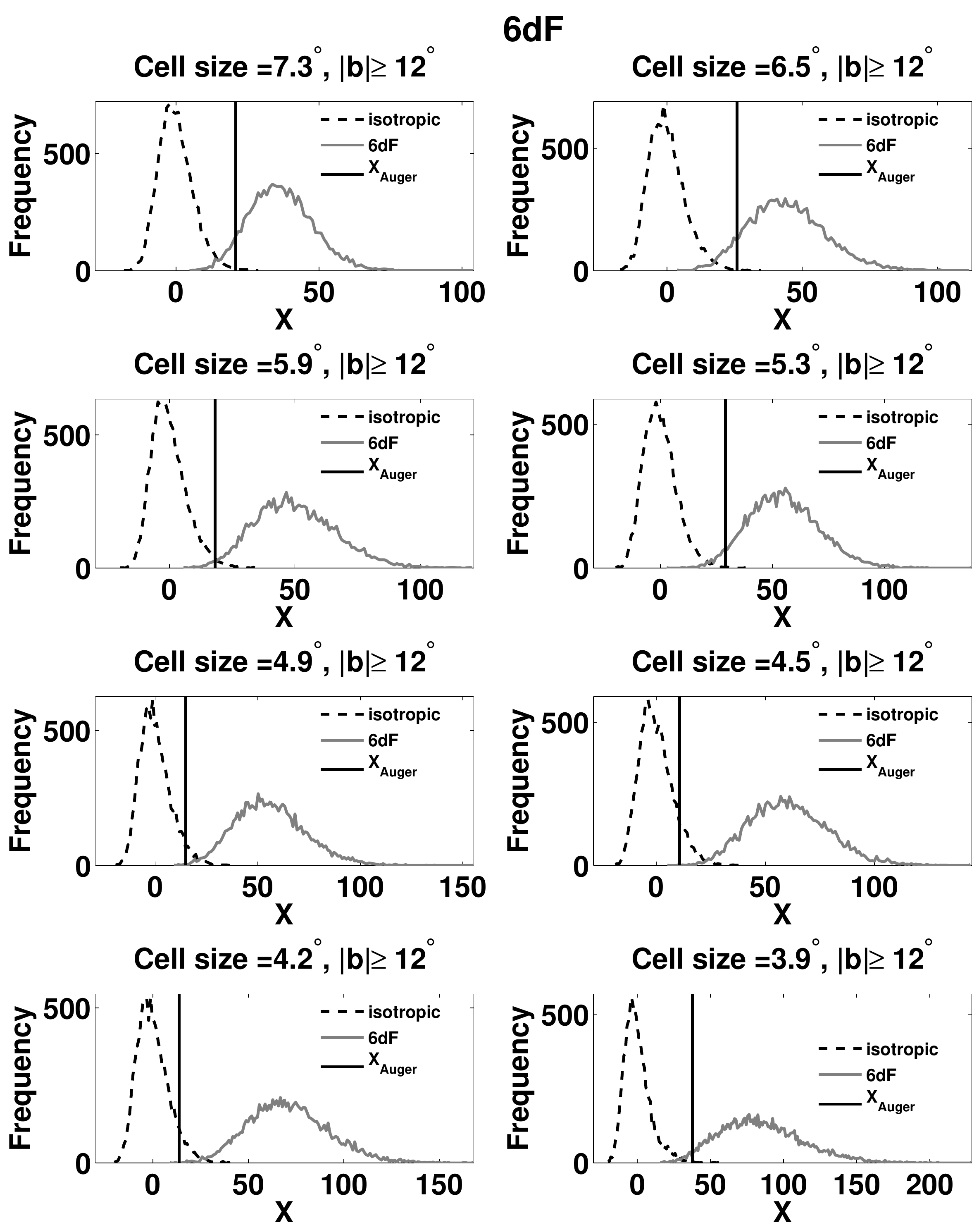}
\caption{The same distributions as in figure \ref{fig:histograms_cell_size_69_pscz} but using the 6dF catalogue.}
\label{fig:histograms_cell_size_69_6df}
\end{figure}
\begin{figure}
\centering
\includegraphics[width=1.20\figlength,  height=0.33\textheight]{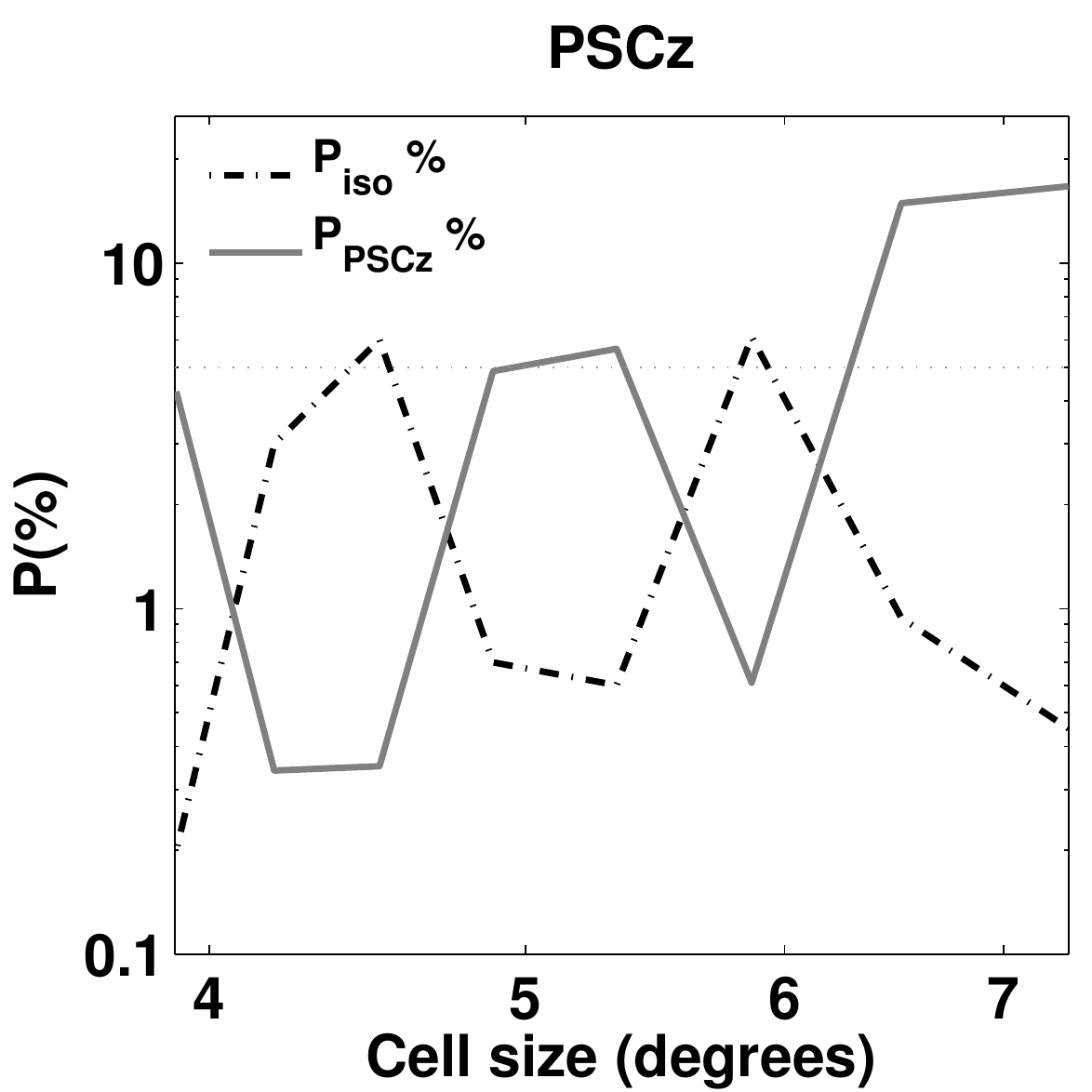}
\includegraphics[width=1.20\figlength,  height=0.33\textheight]{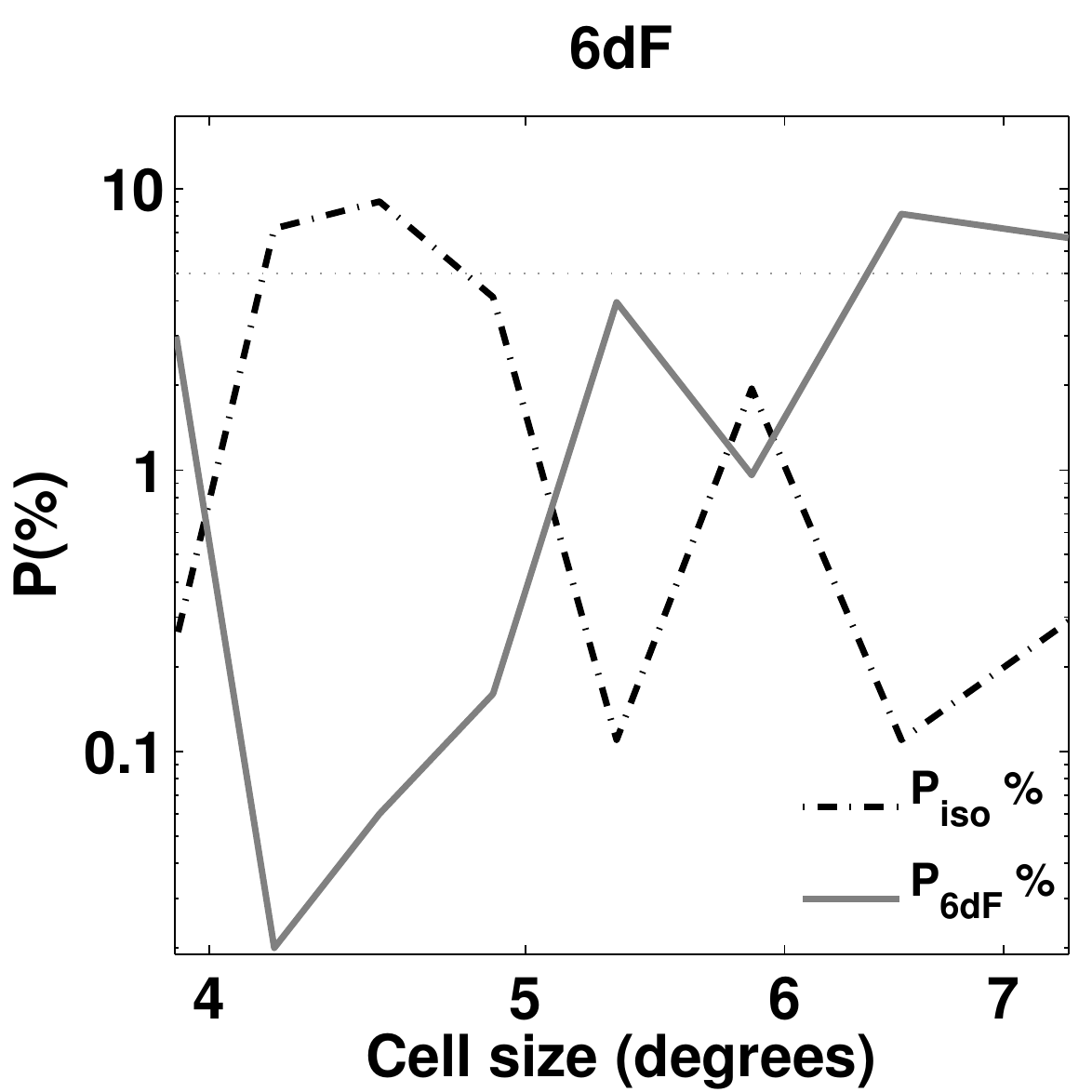}
\caption{\textit{Left:} $P_{\rm{iso}}$ - dashed line ($P_{\rm{PSCz}}$ - grey solid line), the percentage of realisations of UHECR samples drawn from an isotropic (following the PSCz) source distribution in which the value of $X$ \eqref{eq:statistic} obtained was more extreme than $X_{\rm{Auger}}$, the value of $X$ obtained for the observed PAO UHECRs, as a function of cell size. The horizontal dotted line shows the $95\%$ confidence level. \textit{Right:} Same as on the \textit{left} plot but using the 6dF survey.}
\label{fig:p_values_cell_size_dependence}
\end{figure}
\begin{figure}
\centering
\includegraphics[width=1.1\figlength,  height=0.33\textheight]{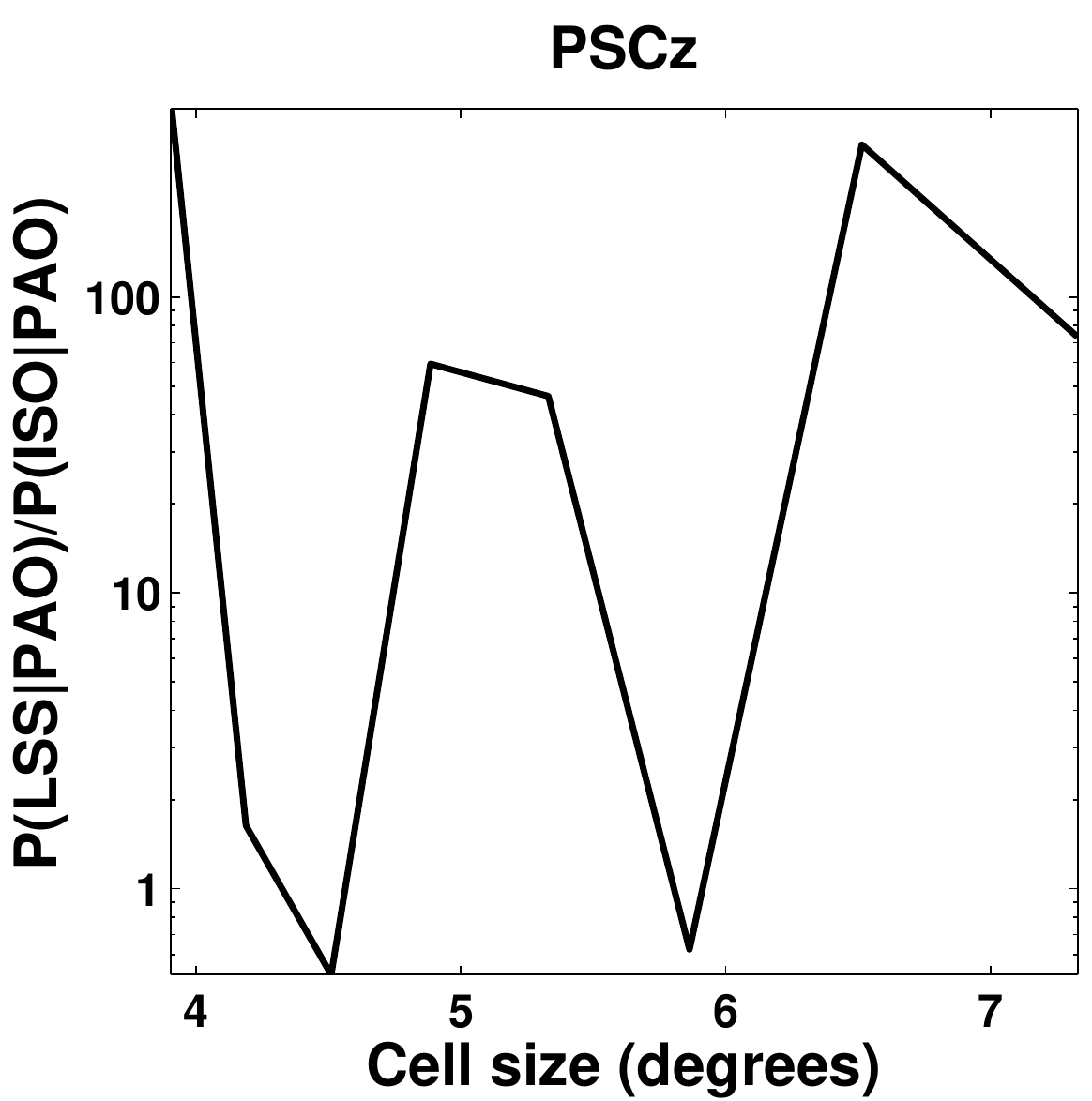}
\includegraphics[width=1.2\figlength,  height=0.33\textheight]{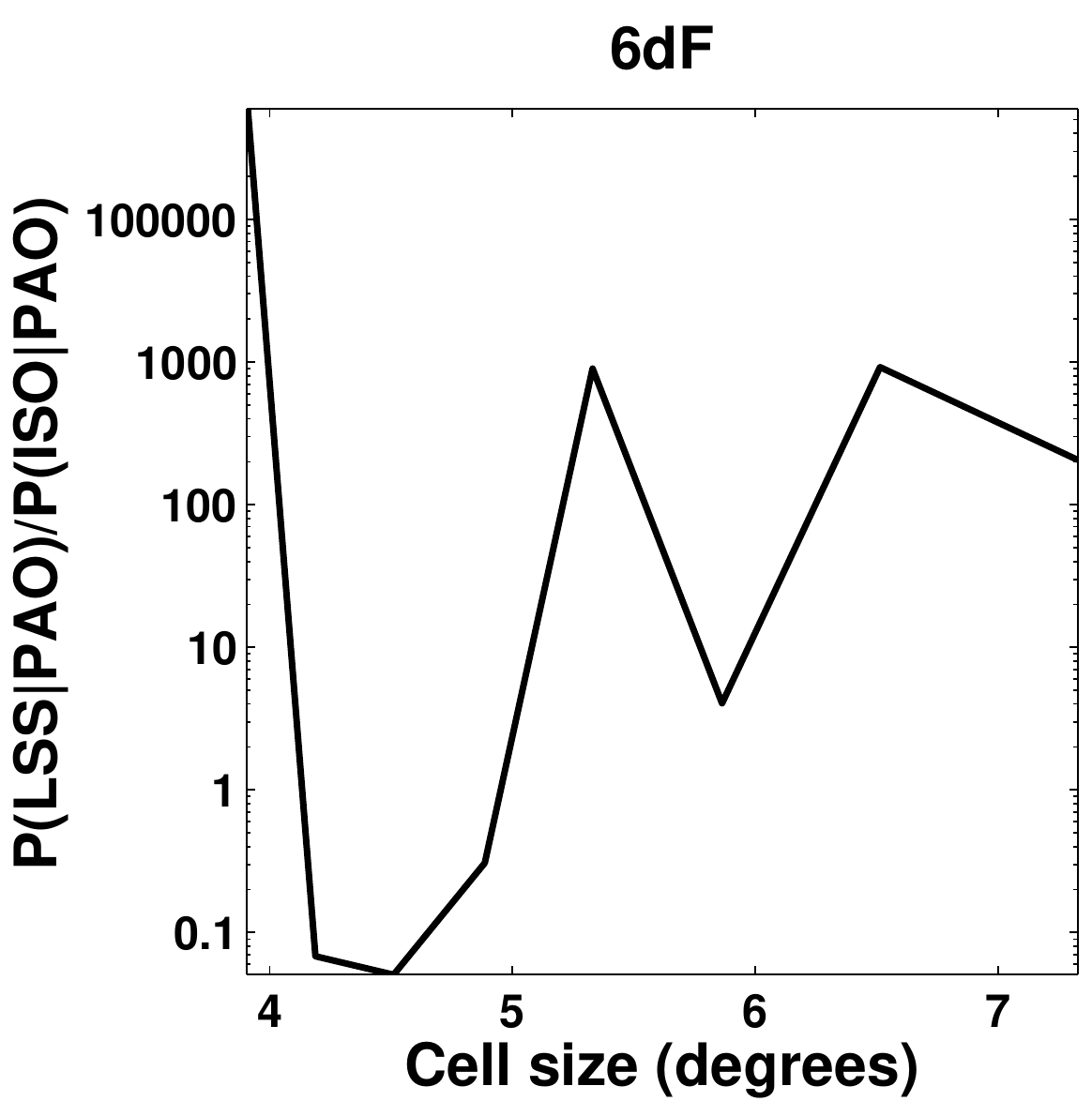}
\caption{\textit{Left:} The quantity $\frac{P(\rm{LSS}|\rm{PAO})}{P(\rm{ISO}|\rm{PAO})}$ \eqref{eq:evidence}, the ratio of the frequencies with which a value of $X$ equal to  $X_{\rm{Auger}}$ (the value of $X$ obtained with the observed PAO UHECRs) is obtained in realisations of UHECR sets from the 2 model source distributions (following the PSCz - \textit{LSS}, isotropic -\textit{ISO}) as a function of cell size. \textit{Right:} Same as on the \textit{left} plot but using the 6dF survey.}
\label{fig:evidence_69_events_cell_size}
\end{figure}
\begin{figure}		
\centering
\includegraphics[width=2.0\figlength,  height=0.2\textheight]{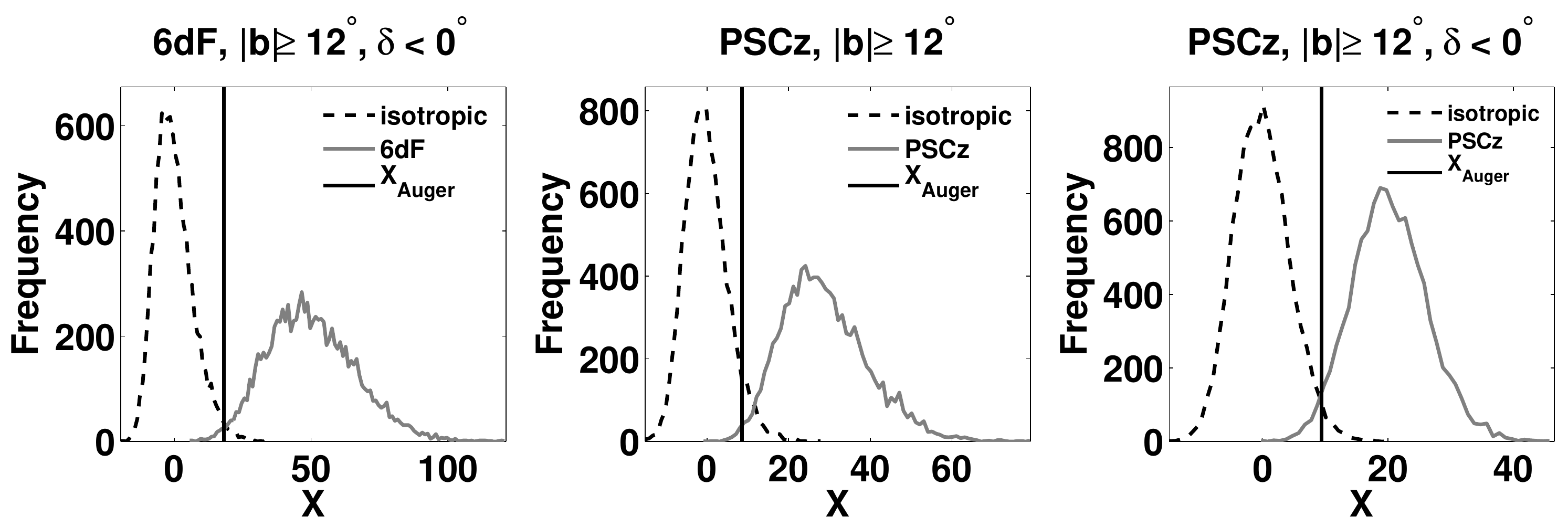}
\caption{The distribution of values of the statistic $X$ \eqref{eq:statistic} obtained in 10000 realisations of UHECRs from an isotropic source distribution (\textit{dashed histograms}), from a distribution of sources that follows LSS (\textit{solid grey histograms}) and $X_{\rm{Auger}}$, the value of $X$ for the observed PAO events (\textit{black solid line}). The distribution of sources that follows LSS was modelled with the 6dF (\textit{left}),  the PSCz (\textit{centre}), and the PSCz sources that lie in the southern hemisphere, i.e. the the PSCz sources in the 6dF field of view (\textit{right}). The plots shown are for cell size $5.9^{\circ} \times 5.9^{\circ}$.}
\label{fig:6df_pscz}
\end{figure}
\begin{figure}
\centering
\subfloat{\includegraphics[width= 2\figlength,  height=0.2\textheight]{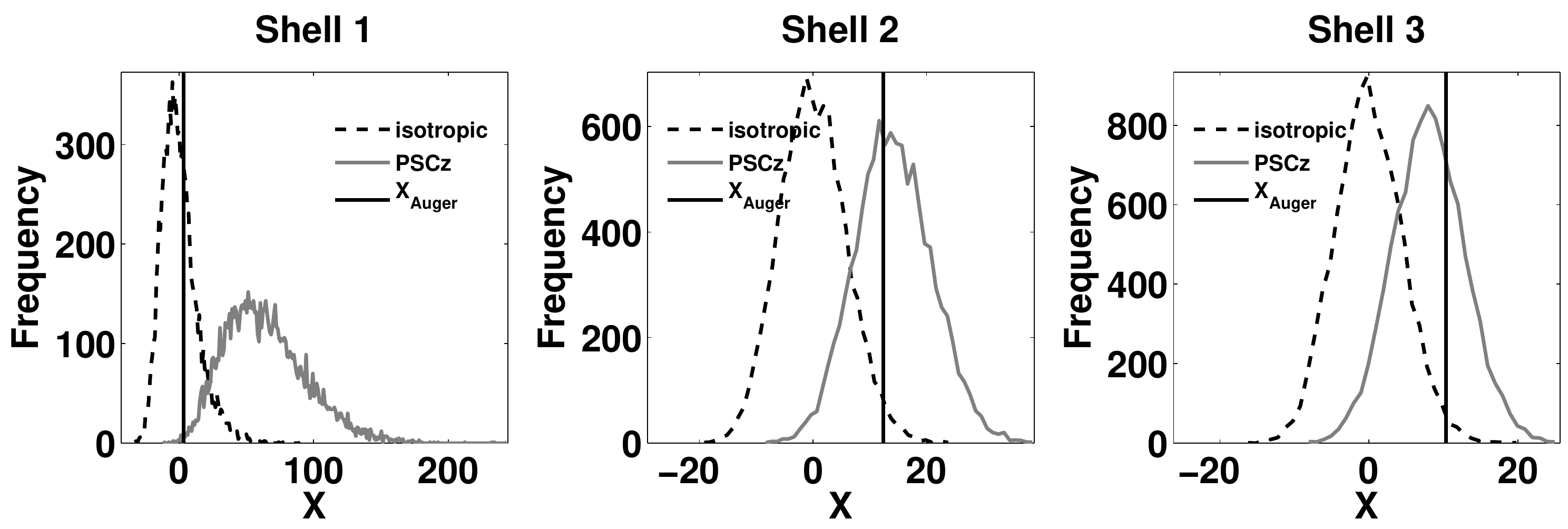}}\\
\subfloat{\includegraphics[width=0.65\figlength,  height=0.2\textheight]{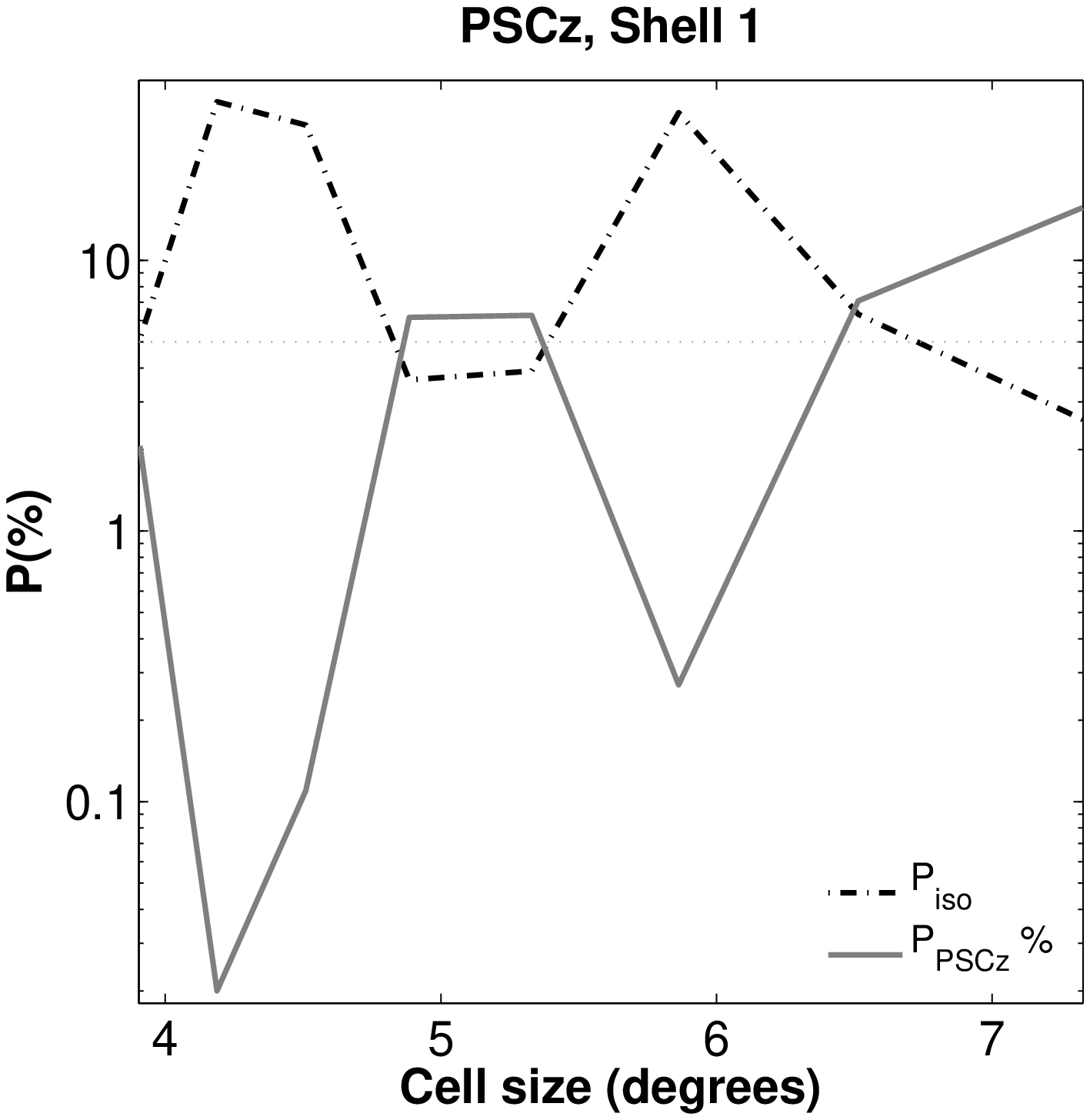}}
\subfloat{\includegraphics[width=0.65\figlength,  height=0.2\textheight]{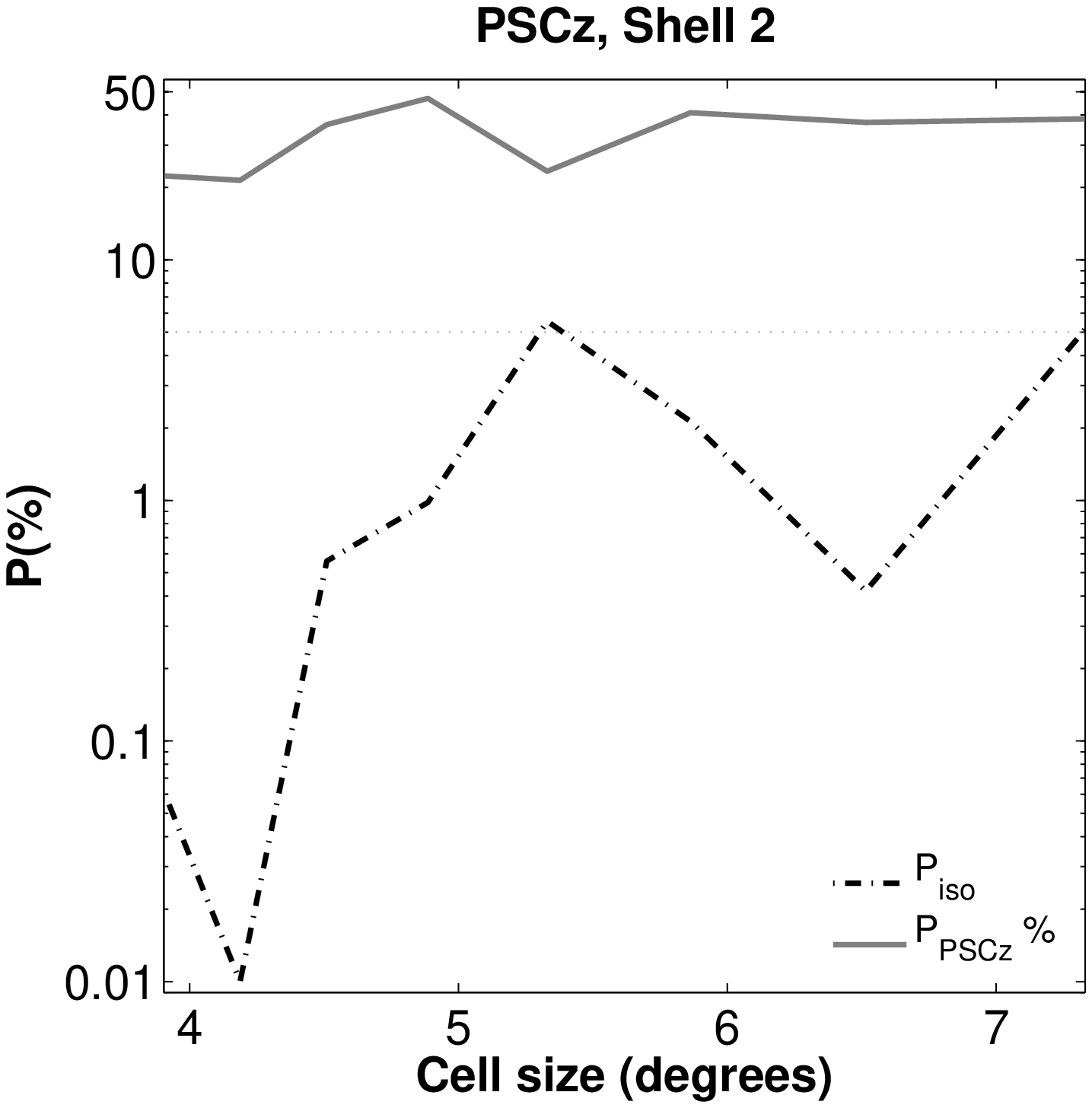}}
\subfloat{\includegraphics[width=0.65\figlength,  height=0.2\textheight]{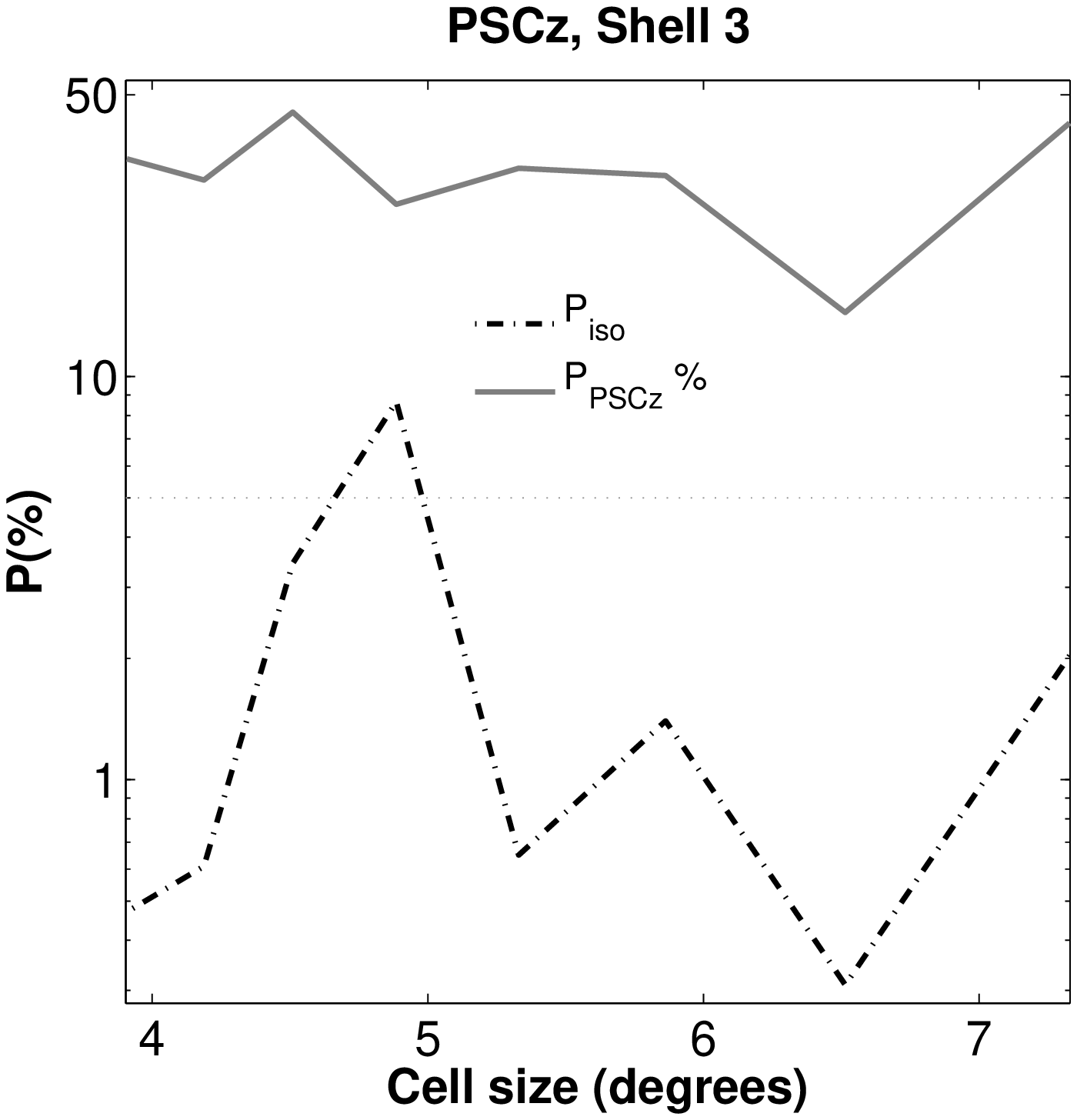}}\\
\subfloat{\includegraphics[width=0.65\figlength,  height=0.2\textheight]{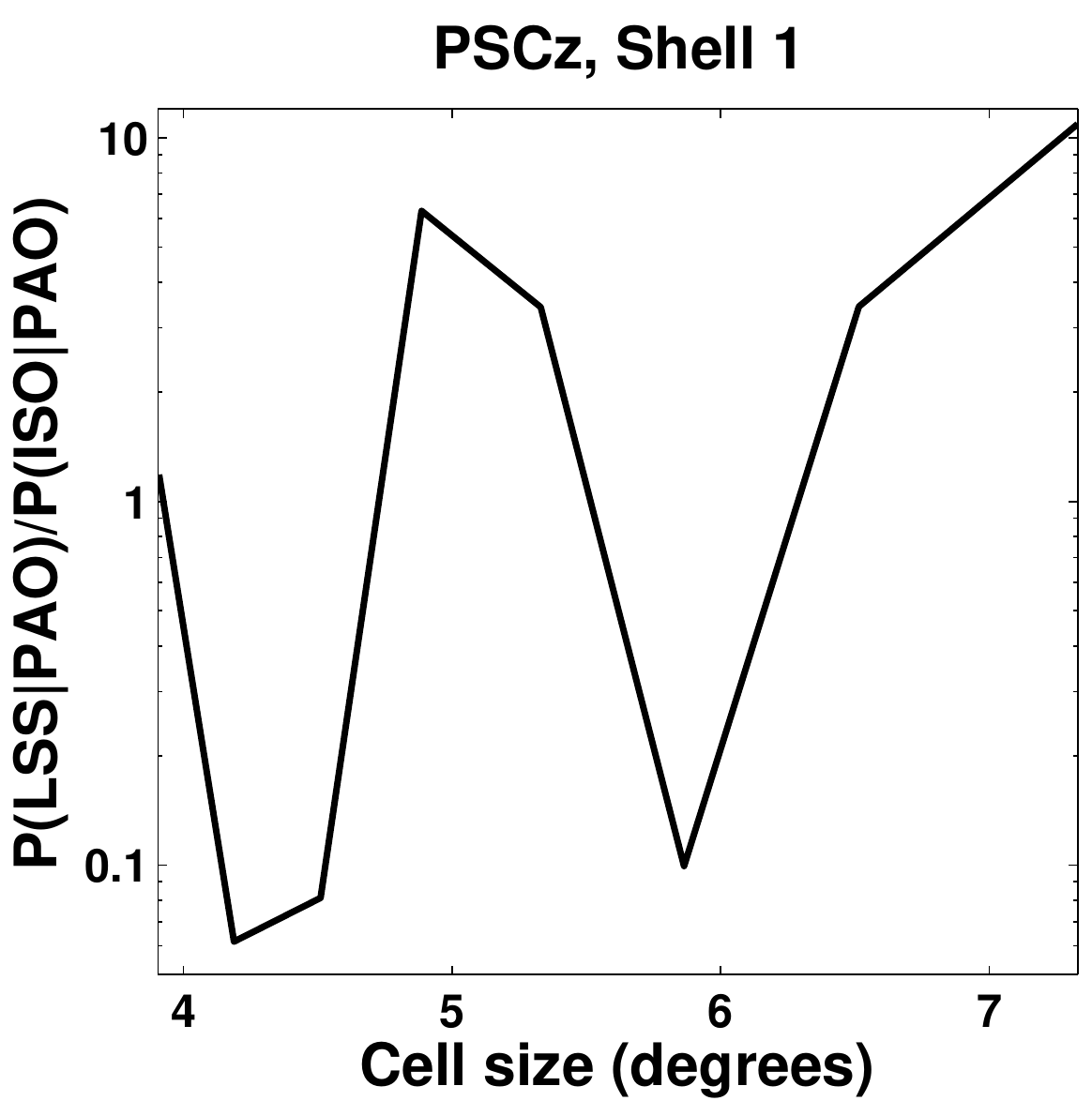}}
\subfloat{\includegraphics[width=0.65\figlength,  height=0.2\textheight]{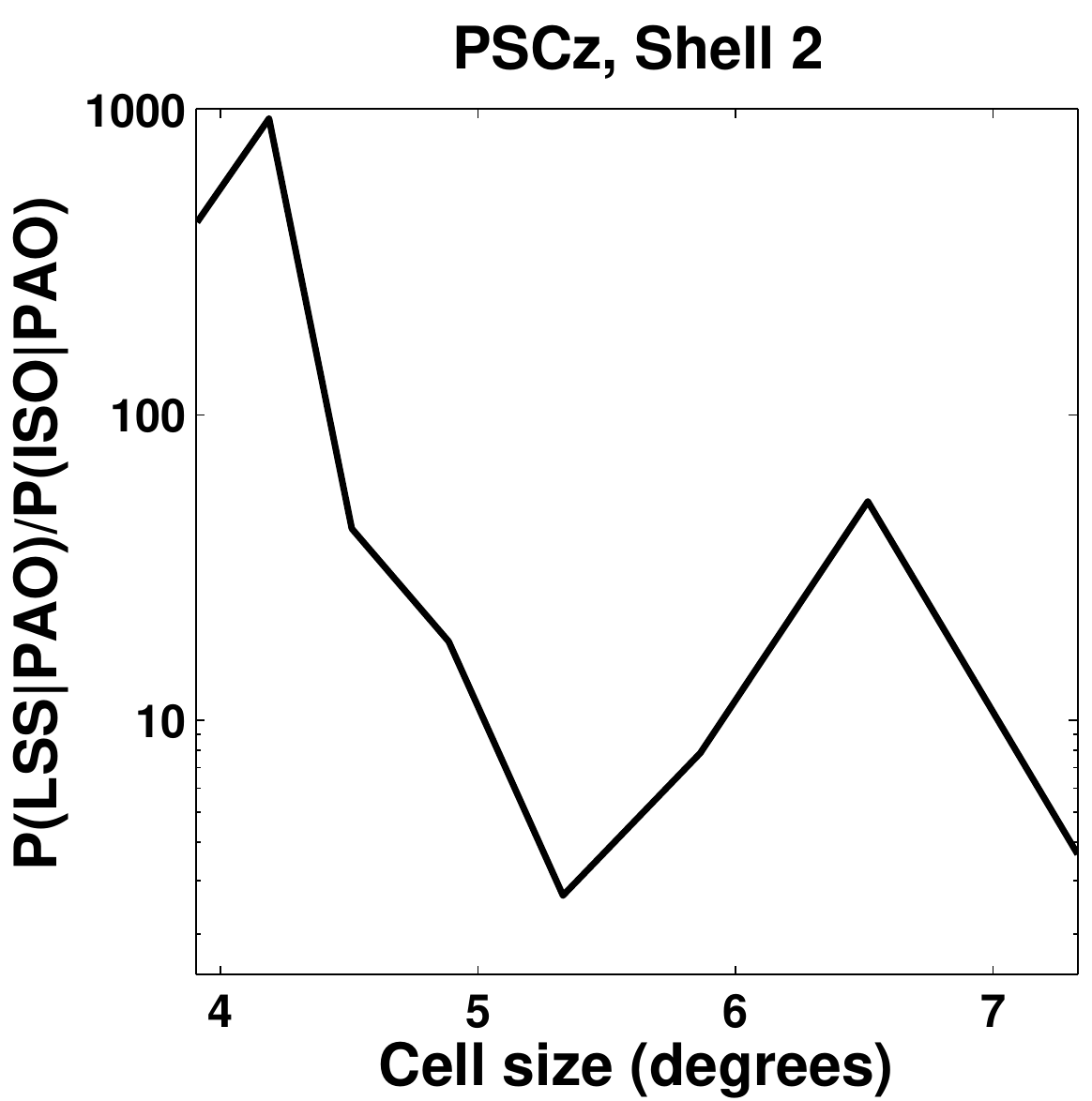}}
\subfloat{\includegraphics[width=0.65\figlength,  height=0.2\textheight]{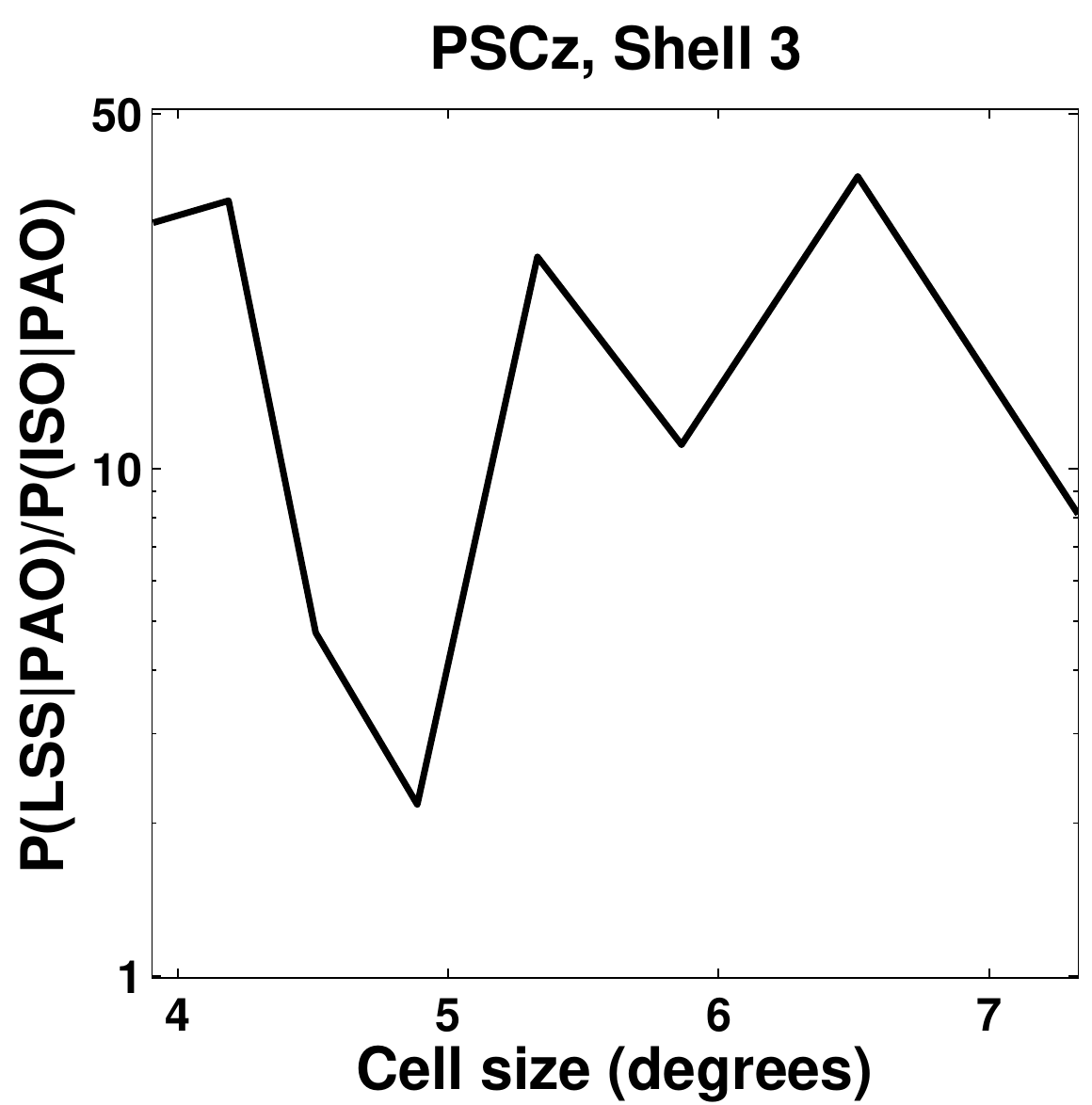}}\\
\caption{\textit{Top Row:} Distribution of values of the statistic $X$ \eqref{eq:statistic} for 10,000 realisations of UHECR sets drawn from an isotropic source distribution (\textit{dashed histograms}), a distribution of sources based on the PSCz (\textit{solid histograms}) and $X_{\rm{Auger}}$, the value of $X$ obtained for the observed PAO UHECRs (\textit{black solid line}) for shells 1, 2 and 3 (see $\S$\ref{subsec:shells}). The cell size used for these plots is $5.9^{\circ} \times 5.9^{\circ}$.\\
\noindent \textit{Middle Row:} $P_{\rm{iso}}$ - \textit{dashed line} ($P_{\rm{PSCz}}$ - \textit{solid line}), the percentage of realisations of sets of UHECRs from an isotropic (correlated with LSS) source distribution in which the value of $X$ was more extreme than $X_{Auger}$ as a function of cell size.\\
\noindent \textit{Bottom Row:} $\frac{P(\rm{PSCz}|\rm{PAO})}{P(\rm{ISO}|\rm{PAO})}$ \eqref{eq:evidence}, the ratio of realisations of the two models that had a value of $X$ equal to $X_{Auger}$ as a function of cell size.}
\label{fig:shells_pscz}
\end{figure}
\begin{figure}
\centering
\subfloat{\includegraphics[width=2\figlength,  height=0.2\textheight]{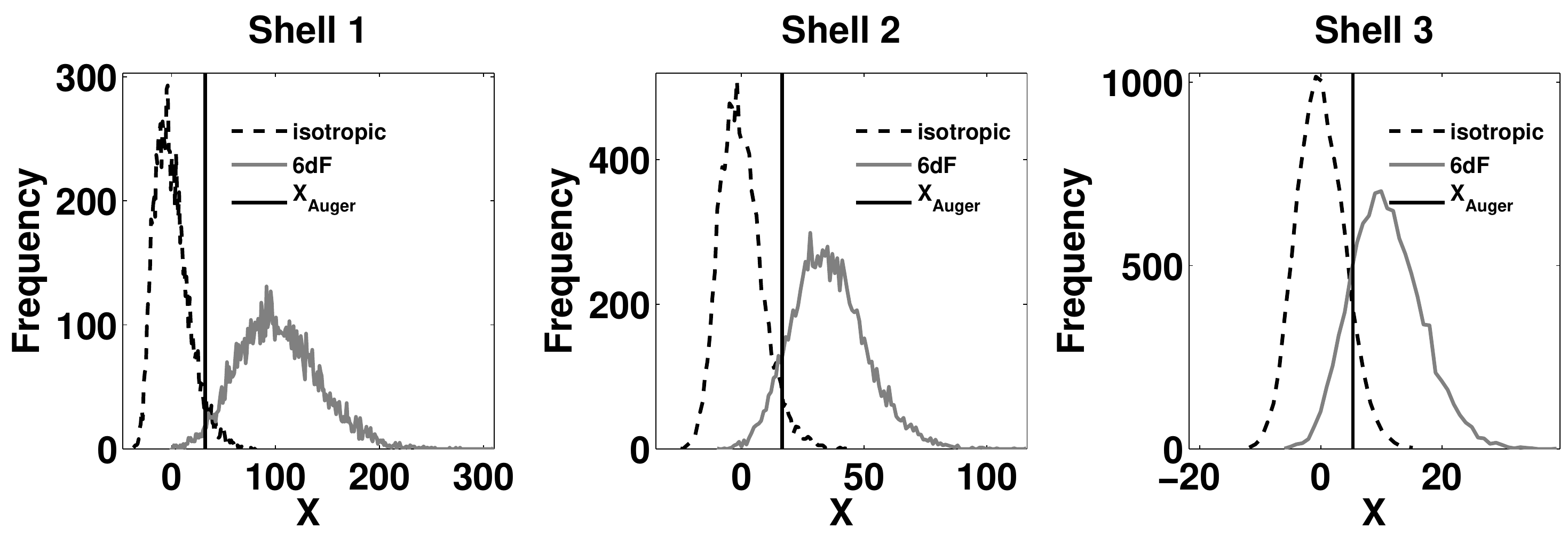}}\\
\subfloat{\includegraphics[width=0.65\figlength,  height=0.2\textheight]{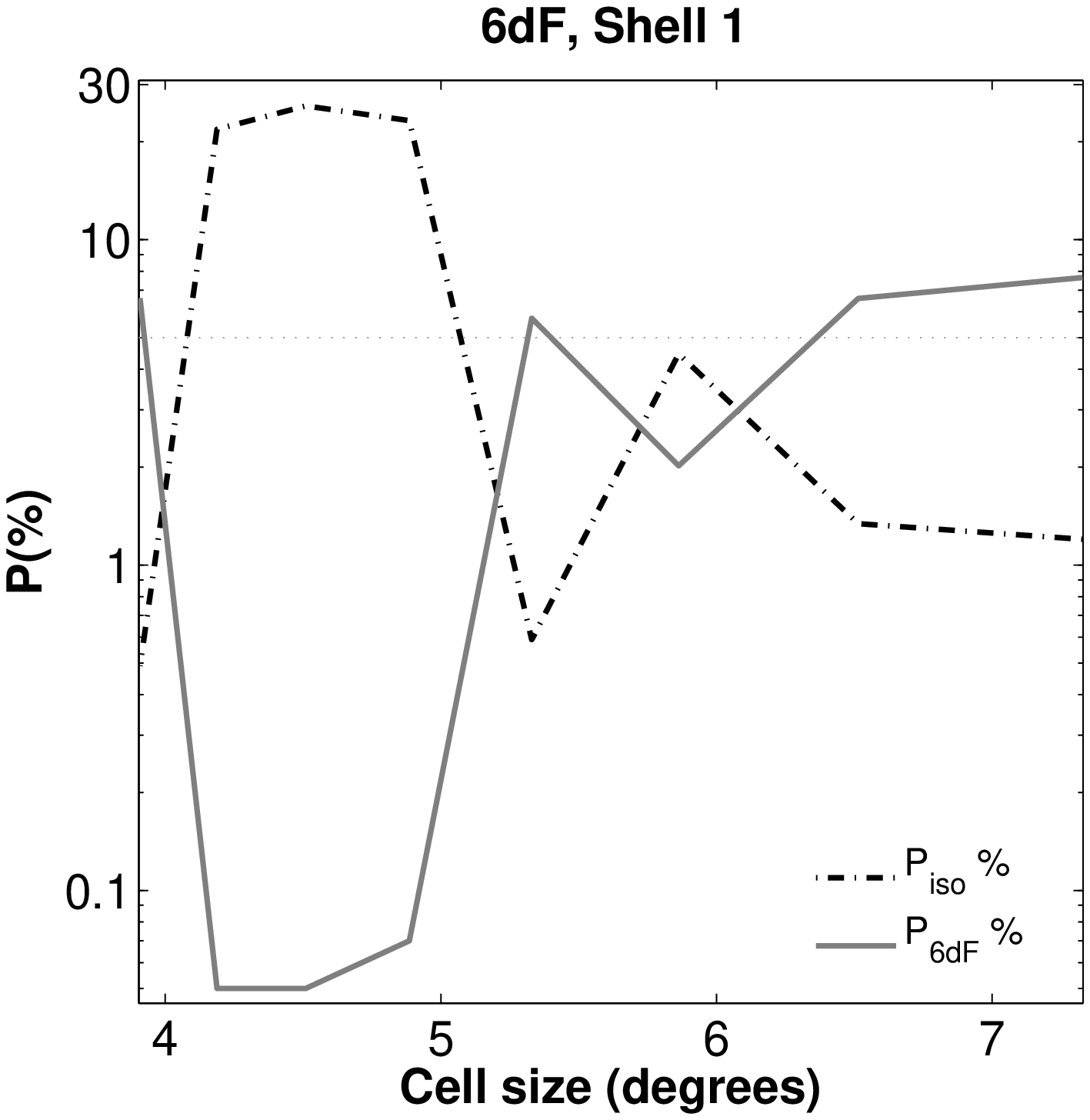}}
\subfloat{\includegraphics[width=0.65\figlength,  height=0.2\textheight]{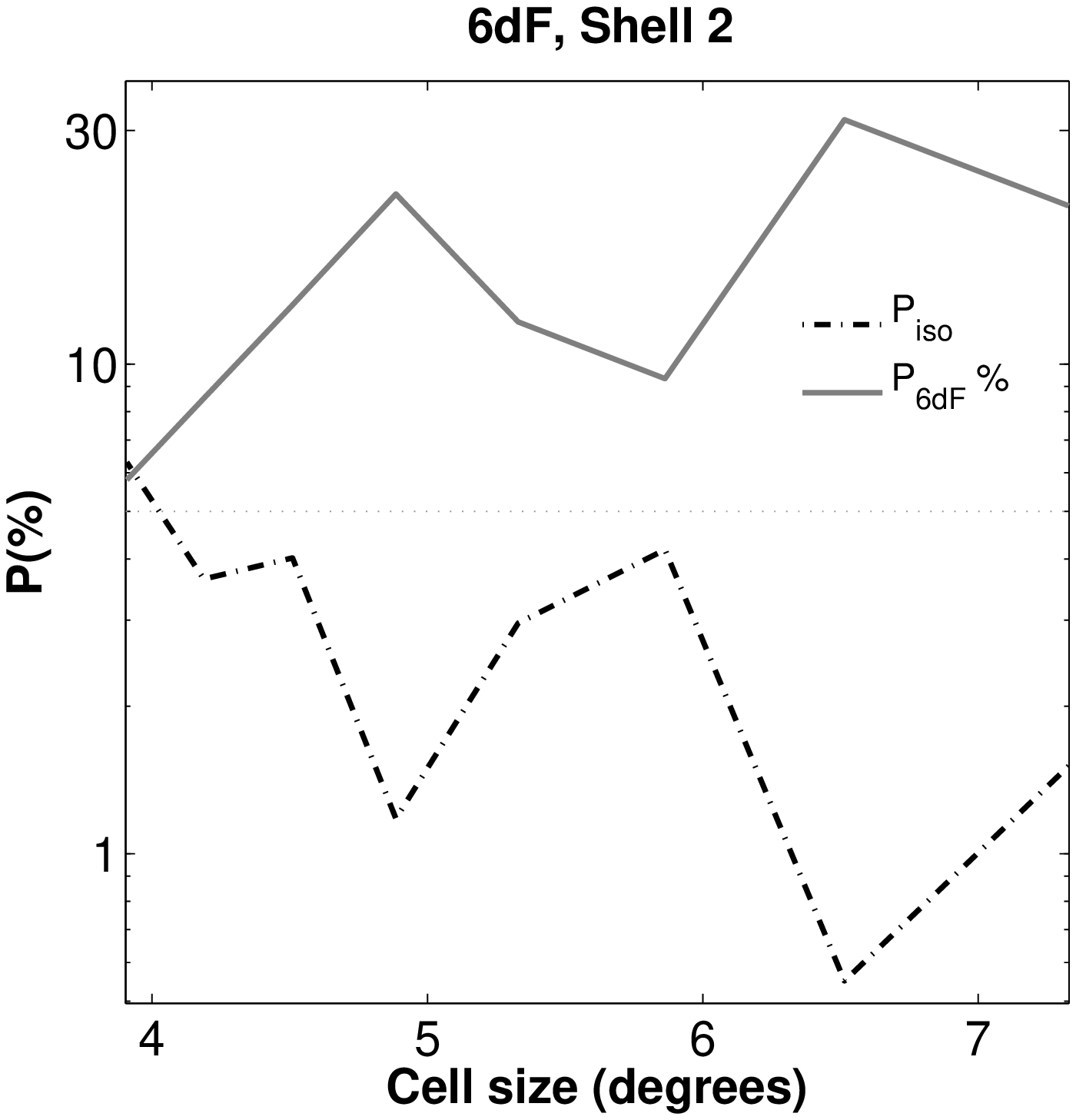}}
\subfloat{\includegraphics[width=0.65\figlength,  height=0.2\textheight]{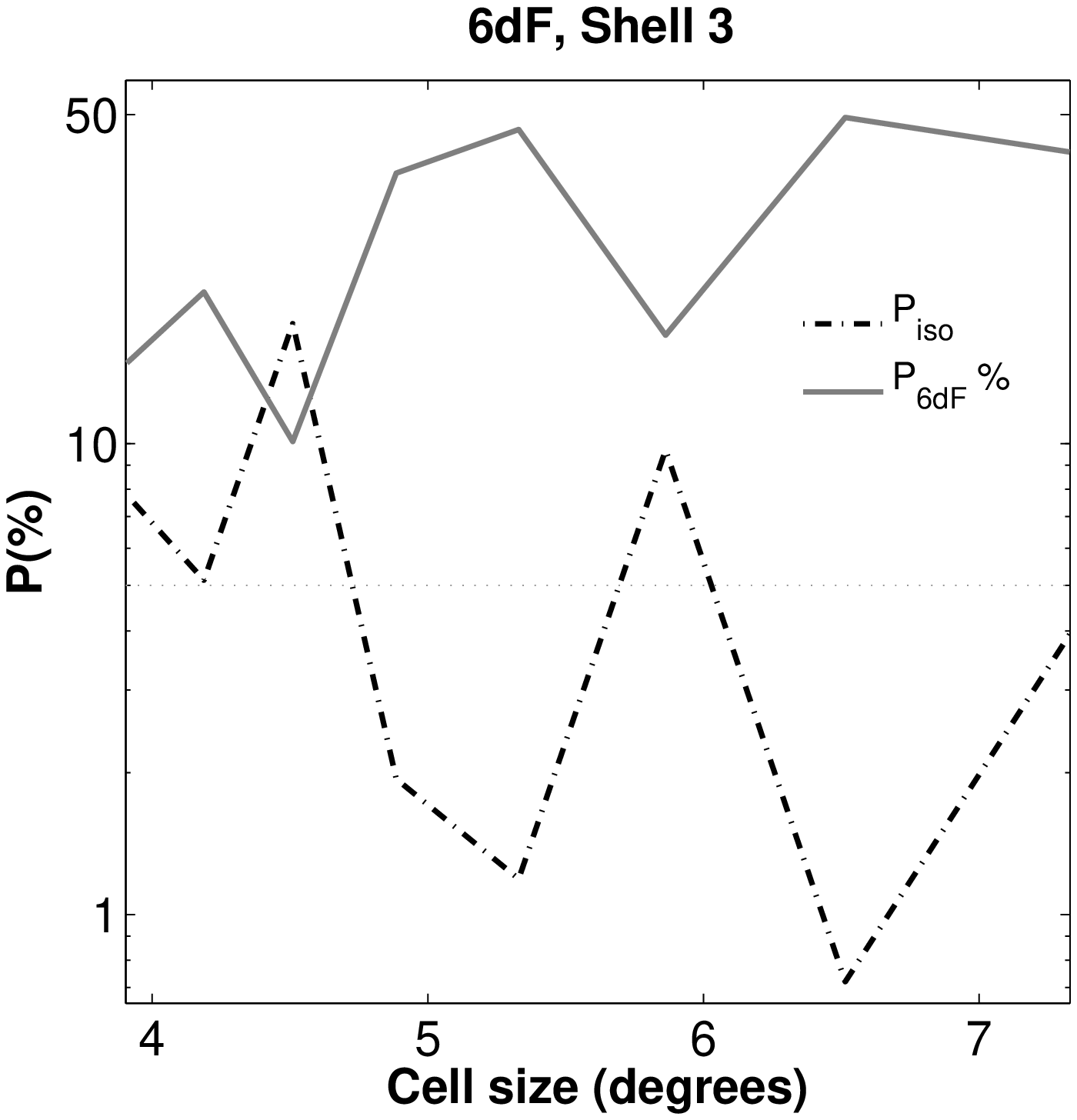}}\\
\subfloat{\includegraphics[width=0.65\figlength,  height=0.2\textheight]{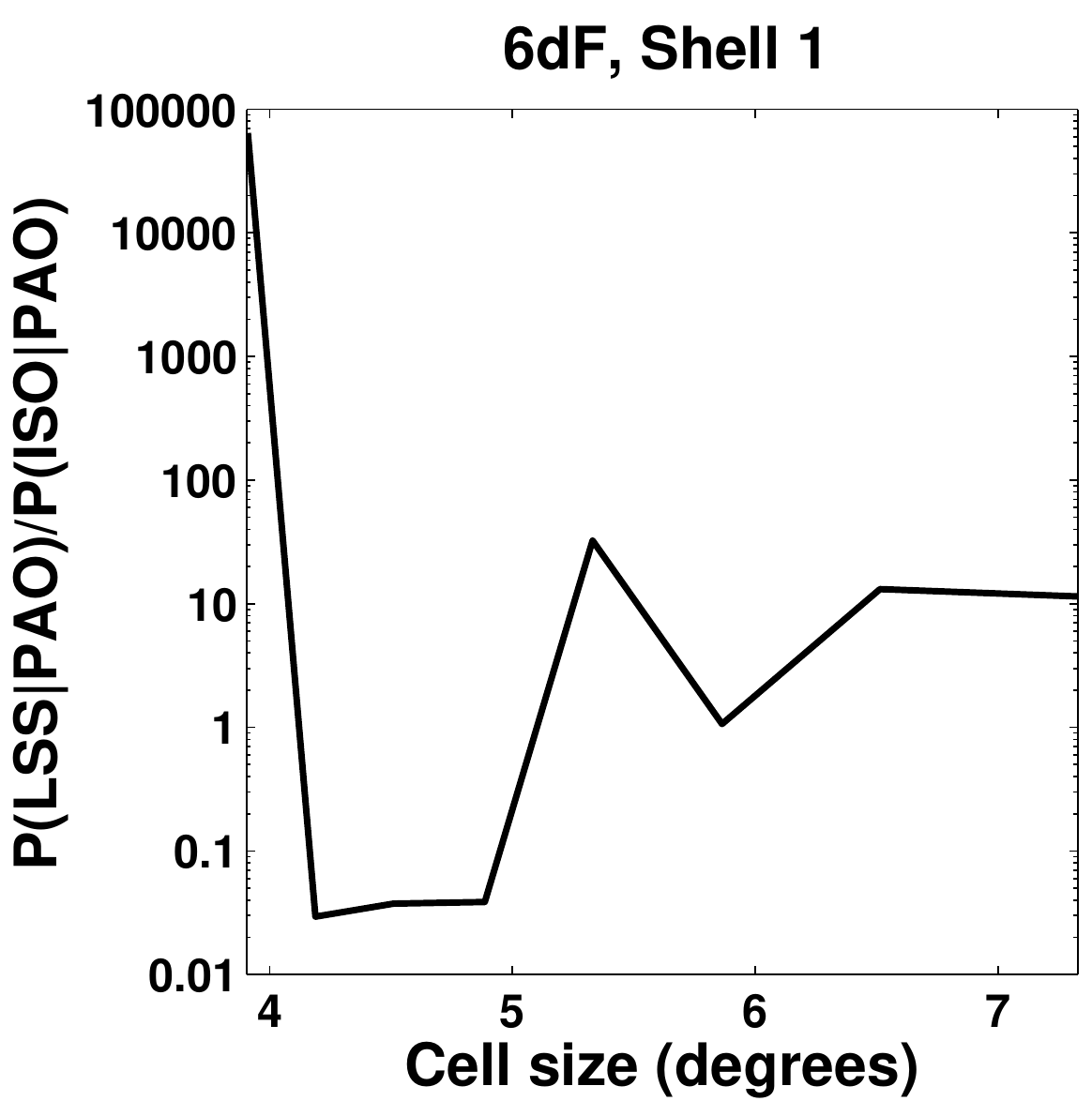}}
\subfloat{\includegraphics[width=0.65\figlength,  height=0.2\textheight]{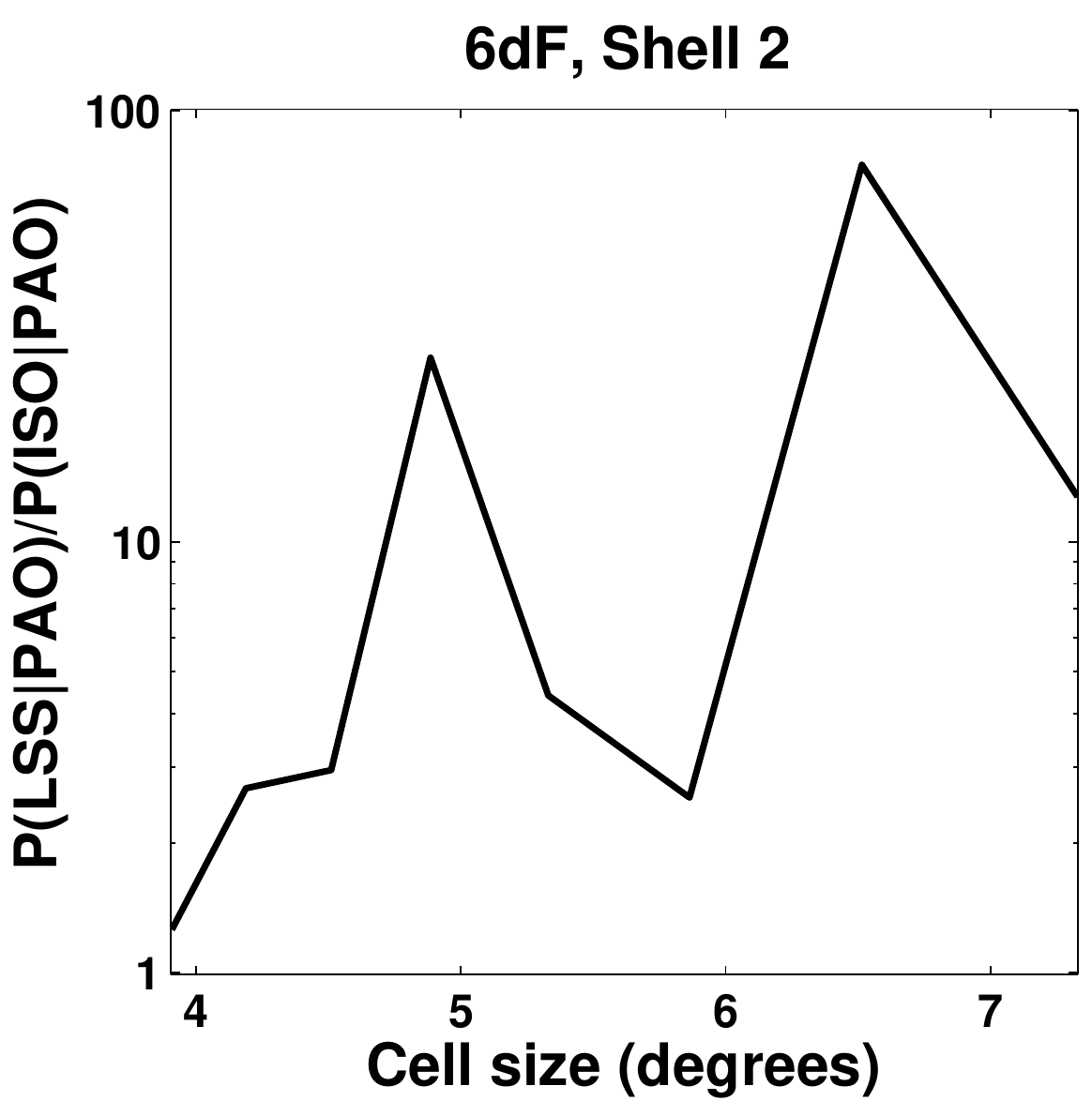}}
\subfloat{\includegraphics[width=0.65\figlength,  height=0.2\textheight]{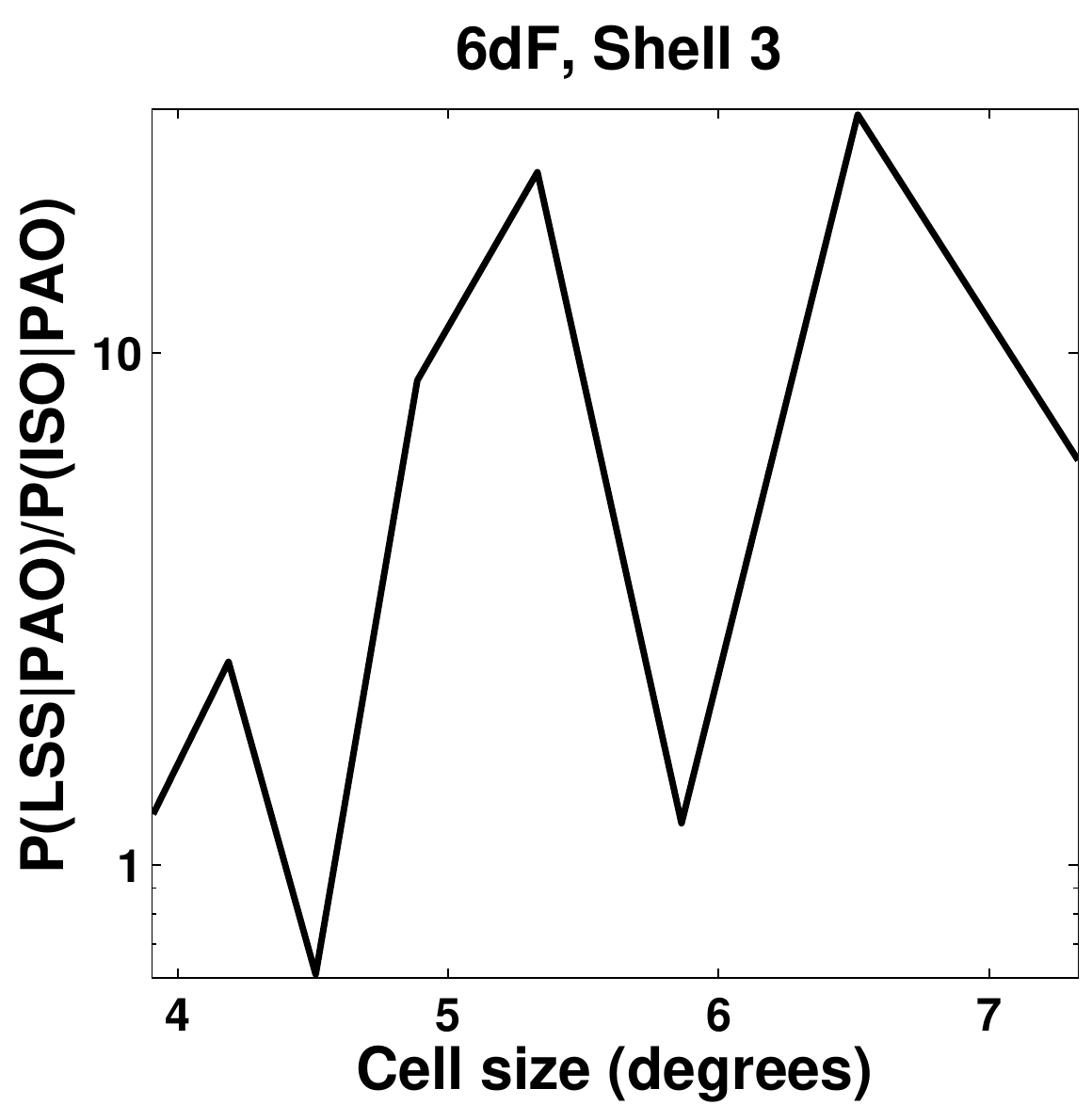}}\\
\caption{Same as in figure \ref{fig:shells_pscz} but for the 6dF survey.}
\label{fig:shells_sixdf}
\end{figure}
\begin{figure}
\centering
\subfloat{\includegraphics[width= 2\figlength,  height=0.2\textheight]{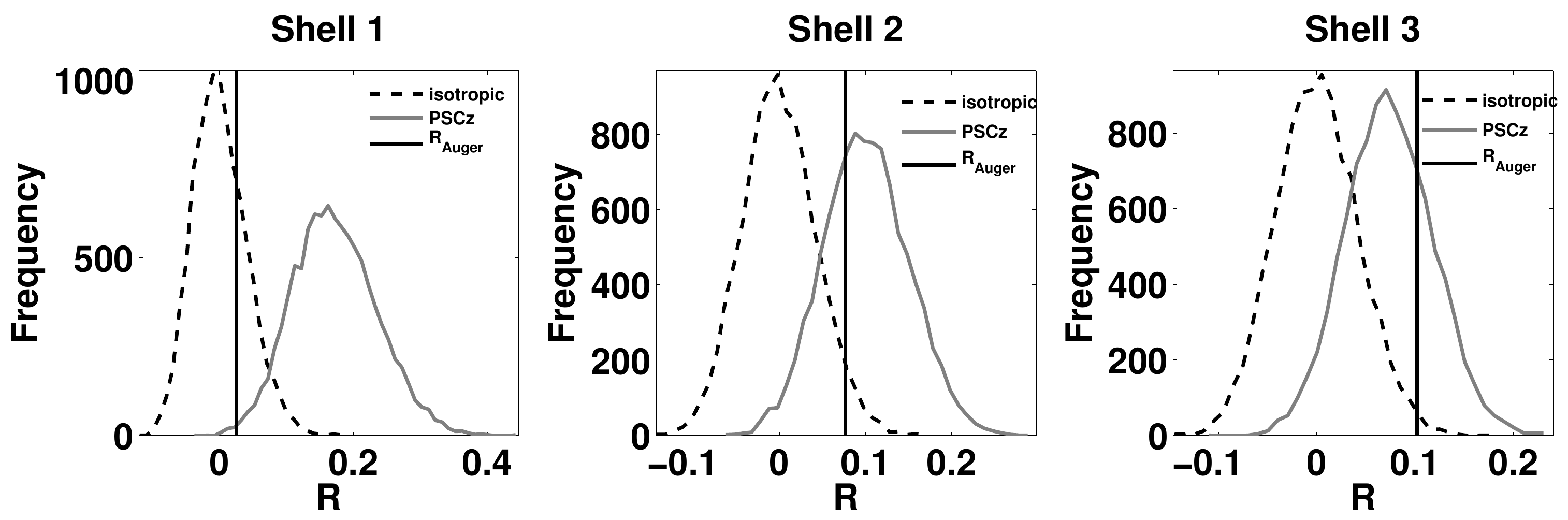}}\\
\subfloat{\includegraphics[width= 2\figlength,  height=0.2\textheight]{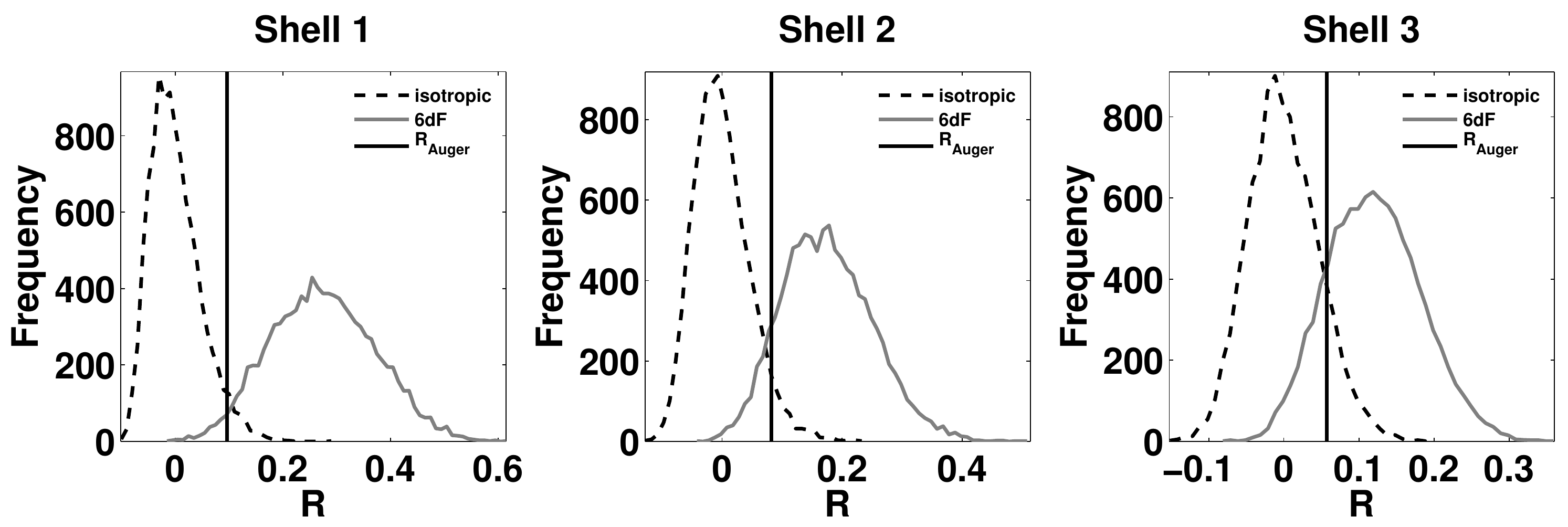}}\\
\caption{\textit{Top Row:} Distribution of values of the normalised correlation coefficient $R$ \eqref{eq:normalised_cross_corr_coeff} for 10,000 realisations of UHECR sets drawn from an isotropic source distribution (\textit{dashed histograms}), from a distribution of sources based on the PSCz (\textit{solid histograms}) and $R_{\rm{Auger}}$, the value of $R$ obtained for the observed PAO UHECRs (\textit{black solid line}), for Shells 1, 2 and 3 (see section \ref{subsec:shells}). The cell size used for these plots is $5.9^{\circ} \times 5.9^{\circ}$. \textit{Bottom Row:} Same as in the row above, but with the 6dF survey.}
\label{fig:normalised_corr_coeff}
\end{figure}
\begin{figure}
\centering
\subfloat{\includegraphics[width=1.3\figlength,  height=0.38\textheight]{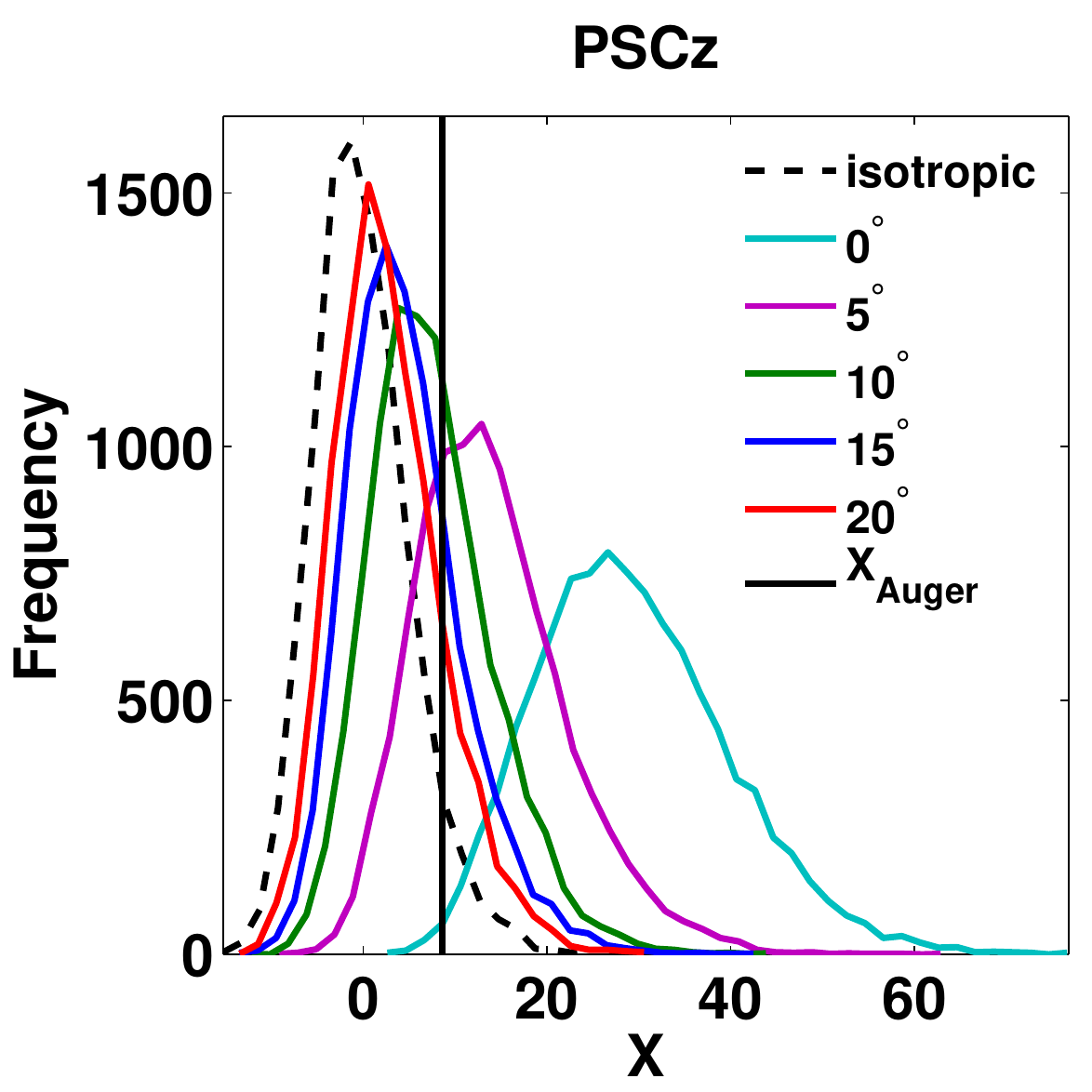}}
\subfloat{\includegraphics[width=1.3\figlength,  height=0.38\textheight]{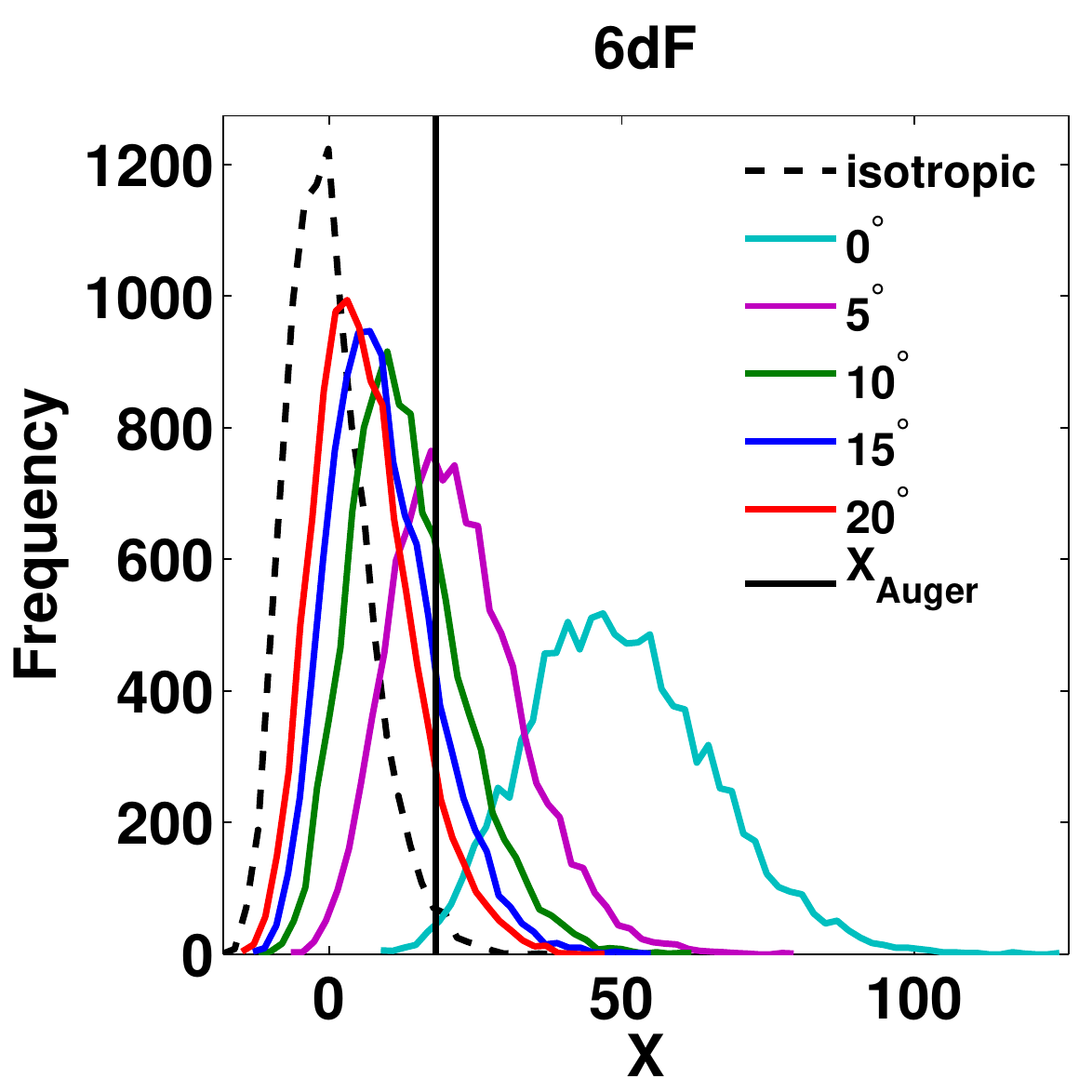}}
\caption{\textit{Left:} The distribution of values of $X$ in 10,000 realisations of UHECRs from an isotropic distribution of sources (\textit{dashed histogram}) and from a distribution of sources based on the PSCz (\textit{solid histograms}) with simulated magnetic deflections. In each solid histogram the UHECRs arrive on Earth with a mean position centred at the source's true position and random angular deflections with amplitude $d$ given in the legend (see $\S$\ref{subsec:deflection}). The cell size shown in this plot is $5.9^{\circ} \times 5.9^{\circ}$. \textit{Right:} Same as on the plot on the \textit{left} but using the 6dF.}
\label{fig:deflection_limit}
\end{figure}
\indent The methodology presented in Section \ref{sec:Methodology} was applied to the 69 UHECRs observed by the PAO with energy $E\geq 55$~EeV until the end of 2009 \cite{2010APh....34..314T}. We perform our analysis by dividing the sky in equal area bins (counts-in-cells) using the HEALPix package \cite{2005ApJ...622..759G}. The disadvantages of the counts-in-cells scheme are the effect of boundaries and the ability to arbitrarily choose the bin size. We deal with both these limitations by treating the size of the equal area bins as a free parameter in a range that covers expected random proton UHECR magnetic deflections ($\S$\ref{subsec:cell_size}). In $\S$\ref{subsec:shells} we study the cross-correlation of the observed PAO UHECRs in shells by dividing the galaxy distribution into redshift shells of equal predicted UHECR flux. In $\S$\ref{subsec:deflection} we relax the assumption of proton UHECRs and investigate the sensitivity of our results to the magnitude of the magnetic deflection. Since the galactic plane and regions of the sky not covered by the galaxy survey are always excluded by our analysis, the number of observed UHECRs that remain in the mask defined region is never 69 and depends on the survey used and the bin size considered. In $\S$\ref{subsec:systematics} we investigate the dependence of our results on systematic uncertainties. 
\subsection{Cross-correlation of UHECRs and nearby LSS}
\label{subsec:cell_size}
In figure \ref{fig:maps} we show the predicted distribution of sources of $55$~EeV UHECRs based on the PSCz/6dF catalogues. To aid with illustration, we plot the model source distribution smeared with a 2 dimensional Gaussian filter with standard deviation $\sigma = 7.2^{\circ}$, whereas throughout the rest of this work we have treated galaxies as discrete point sources. In the maps of the PSCz one can see a very large contribution from the Virgo cluster $(l \sim -80^{\circ}, b \sim 75^{\circ})$, Hydra-Centaurus  $(-60^{\circ} \leq l \leq 0^{\circ}, 0^{\circ} \leq b \leq 45^{\circ})$ and the Perseus-Pisces  supercluster $(l \sim140^{\circ}, b \sim -25^{\circ})$. In the maps of the 6dF there is a large excess as a result of flux from Hydra-Centaurus and the Shapley Concentration (centred at $(l \sim -50^{\circ}, b \sim 30^{\circ})$ at a distance $\sim 200$~Mpc). We plot the source distribution with those of the 69 observed PAO UHECRs in the mask defined region superimposed in figure \ref{fig:maps_and_UHECRs}. Note that we haven't masked with the PAO exposure in these plots. After doing so the relative weights of structures change so that for example Virgo no longer dominates in the PSCz derived model.\\
\indent Because the observed galaxy distribution is in the form of flux limited as opposed to volume limited samples, we have weighted observed galaxies in the PSCz and 6dF with the inverse of the selection function $\psi(r_c)$ ($\S$\ref{subsec:gzk_weight}). There is some uncertainty associated with this process, however this procedure is well motivated given that matter in the universe is clustered, hence unobserved galaxies that are below the galaxy survey's magnitude limit are more likely to reside near the observed galaxies than in regions where no sources were observed. \\
\indent In figures \ref{fig:histograms_cell_size_69_pscz} and \ref{fig:histograms_cell_size_69_6df} we show the distribution of values that the statistic $X$ takes in 10,000 mock realisations of isotropically distributed UHECRs, as observed by the PAO (dashed histograms) and  the distribution of values of $X$ in 10,000 mock realisations of 55 EeV UHECR protons with sources drawn from the predicted UHECR source distribution that follows the distribution of matter in the PSCz and 6dF catalogues respectively (solid histograms). The black vertical line in each subplot shows the value of $X$ for the observed PAO UHECRs assuming $E_f = 55$~EeV. Each subplot corresponds to a different bin-size in the range $7.3^{\circ} - 3.9^{\circ}$. The picture that emerges from figures \ref{fig:histograms_cell_size_69_pscz} and  \ref{fig:histograms_cell_size_69_6df} is that the observed PAO UHECRs are not consistent with the mean of either one of the two models although there is weak evidence for a source distribution correlated with nearby galaxies in bin sizes $6.5^{\circ} \times 6.5^{\circ}$ and $7.3^{\circ} \times 7.3^{\circ}$.\\
\indent We quantify the significance of this result in figure \ref{fig:p_values_cell_size_dependence}, where we plot the percentage of realisations in which the value of $X$ was more extreme than $X_{\rm{Auger}}$, the value of $X$ obtained with the observed PAO UHECRs, in each of the models of UHECR source distribution considered ($P_{\rm{iso}}$ in the isotropic model and $P_{\rm{PSCz}}$/$P_{\rm{6dF}}$ in the model where the UHECR sources are PSCz/6dF galaxies) as a function of cell size. One first notices that $P_{\rm{iso}}$ and $P_{\rm{survey}}$ are very sensitive to bin size and that this behaviour is not monotonic as a function of bin-size. This is a result of very low UHECR counts making the results susceptible to boundary effects. Overall there is very good agreement between the results obtained with the two galaxy surveys. Although the 6dF and PSCz samples are selected differently (near-infrared and far-infrared respectively) and the 6dF is deeper than the PSCz, they probe approximately the same large scale structure. For most of the parameter space considered the observed PAO UHECRs are inconsistent with isotropy at a level $ \geq 95\%$, as can be seen by the dotted line that gives the $95\%$ confidence level. At the same time they have a lower value of $X$ than $\geq 85\%$ of mock realisations from a source distribution following LSS, which is lower than we would have expected on average if the observed UHECRs originate in galaxies in either survey.\\
\indent Another way to quantify the significance of the results of figures \ref{fig:histograms_cell_size_69_pscz} and \ref{fig:histograms_cell_size_69_6df} is to consider the frequency with which $X$ equal to $X_{\rm{Auger}}$ occurs in mock realisations of the two different models of UHECR source distribution considered, which is proportional to the likelihood ratio of the two models. If this ratio is greater than 1 then the model in the numerator is preferred by the observed data, whereas if it is smaller than one the model in the denominator is preferred. We show this ratio 
\begin{equation}
\frac{P(\textrm{LSS source model is true}|X_{\rm{Auger}})}{P(\textrm{isotropic source model is true}|X_{\rm{Auger}})},
\end{equation} 
\noindent hereafter 
\begin{equation}
\frac{P(\rm{LSS}|\rm{PAO})}{P(\rm{ISO}|\rm{PAO})}
\label{eq:evidence}
\end{equation} 
\noindent separately for the two galaxy catalogues as a function of cell size in figure \ref{fig:evidence_69_events_cell_size}. Here again we see that the results are extremely sensitive to the choice of bin size and much less sensitive to the choice of galaxy survey. A future larger UHECR dataset will suffer significantly less from such fluctuations and allow to draw firmer conclusions within the framework we have presented here. \\
\indent In figure \ref{fig:6df_pscz} we show the results of the correlation analysis performed using the 6dF survey next to the same results for the PSCz, as well as the results of the analysis performed with the PSCz survey but restricted to the Southern hemisphere for cell size $5.9^{\circ} \times 5.9^{\circ}$. This allows for a direct comparison between the results of the two surveys which we confine to the same field of view. Again we see very good agreement between the two surveys.
\subsection{Cross-correlation in equal predicted flux radial shells}
\label{subsec:shells}
In this section we present our proposed new method of studying any correlation between observed UHECR arrival directions and predicted UHECR source distribution by dividing the predicted source distribution into radial shells with distance. If a correlation exists this method will help localise the source population and constrain some of its properties, such as for example the redshift evolution. \\
\indent We divide the expected UHECR source distribution into three shells, each contributing a third of the expected UHECR flux based on the predicted source distribution presented in Section \ref{sec:Methodology}, and cross-correlate the observed PAO UHECRs with the predicted source distribution in each of the shells. In order to define shells contributing equal predicted UHECR flux we determine $r_1$ and $r_2$, such that for each shell:
\begin{equation}
\begin{split}
& \int_{r_{\rm{min}}}^{r_1} \! \frac{d}{dr_L} \left( n(r_L) \cdot \omega_{\rm{gal}}(r_L) \right) \cdot dr_{L} = \int_{r_{1}}^{r_2} \! \frac{d}{dr_L} \left( n(r_L) \cdot \omega_{\rm{gal}}(r_L) \right) \cdot dr_{L}= \\ 
& \indent \int_{r_{2}}^{r_{\rm{max}}} \! \frac{d}{dr_L} \left( n(r_L) \cdot \omega_{\rm{gal}}(r_L) \right) \cdot dr_{L} = \frac{1}{3}\int_{r_{\rm{min}}}^{r_{\rm{max}}} \! \frac{d}{dr_L} \left( n(r_L) \cdot \omega_{\rm{gal}}(r_L) \right) \cdot dr_{L}, 
\end{split}
\end{equation}
\noindent where $n(r_L)$ is the number of sources at distance $r_L$, $\omega_{\rm{gal}}$ is the weighted contribution of a source at distance $r_L$ to the expected UHECR flux defined in \eqref{eq:gal_weight} and $r_{\rm{min}}$ and $r_{\rm{max}}$ are the distances of the nearest and most distant sources in the galaxy survey (in practice $r_{\rm{max}}$ is a cut we apply, well beyond the GZK-horizon, in order to avoid diverging weights of the few, very distant sources, present in the galaxy surveys). Since each nearby galaxy contributes more of the expected UHECR flux than any other more distant galaxy, due to flux suppression with distance and particle energy losses during propagation, these three shells are not equal in width nor in the number of galaxies they contain. In table \ref{tab:shells} we give the number of sources in each of the shells as well as the distances covered by each shell for the PSCz and the 6dF. \\
\begin{table}
\begin{center}
  \begin{tabular}{| l | c | c | c | c | c | c | }
  \hline
 &\multicolumn{3}{c |}{PSCz} &\multicolumn{3}{c |}{6dF} \\ 
 \hline
  & shell 1 & shell 2 & shell 3 & shell 1 & shell 2 & shell 3 \\
  $r_{\rm{start}} - r_{\rm{end}}$~[Mpc] & 0 - 47 & 47 - 132 & 132 - 365 & 0 - 29 & 29 - 100 & 100 - 365 \\
  Number of sources & 2189 & 5577 & 5300 & 1103 & 10711 & 65785 \\
  \hline
  \end{tabular}
  \caption{The number of galaxies and the distances covered in each of the three radial shells ($\S$\ref{subsec:shells}) expected to contribute equal UHECR flux, for a model UHECR source distribution based on the PSCz (\textit{left}) and on the 6dF (\textit{right}). Beyond $\sim100$~Mpc the PSCz selection function, $\psi(r_c)$ \eqref{eq:generic_sel_fn}, drops off quicker than $\omega(r_L)_{\rm{flux}}$ \eqref{eq:weight}, the flux weight for a source at distance $r_L$. As a result the effective weight of each galaxy, $\omega_{\rm{gal}}$ \eqref{eq:gal_weight}, does not monotonically decrease with distance for the PSCz which is why shell 3 contains fewer galaxies than shell 2 (see also $\S$ \ref{subsec:systematics} for a more detailed discussion of the PSCz selection function). }
 \label{tab:shells}
 \end{center}
 \end{table}
\indent In the top row of figures \ref{fig:shells_pscz} and \ref{fig:shells_sixdf} we show the distribution of the values that the statistic $X$ \eqref{eq:statistic} takes in 10000 realisations of UHECR samples from an isotropic distribution of sources and from a distribution of sources based on the local galaxy distribution as well as for the observed PAO UHECRs, in each of the three shells, for $5.9^{\circ} \times 5.9^{\circ}$ bins, for the PSCz and 6dF respectively. In the middle and bottom row we show $P_{\rm{iso}}$, $P_{\rm{PSCz}}$ and $\frac{P(\rm{LSS}|\rm{PAO})}{P(\rm{ISO}|\rm{PAO})}$ (defined in $\S$\ref{subsec:cell_size}) for shells 1, 2 and 3 from left to right, as a function of cell size. We see a strong sensitivity to the choice of cell-size as in $\S$\ref{subsec:cell_size} in these figures. Overall when cross-correlated with shell 1 of the PSCz and shells 1, 2 of the 6dF the observed PAO UHECRs are consistent with isotropy, whereas there is weak evidence for correlation of the observed UHECRs with the predicted source distribution is shells 2 and 3 of the PSCz and with shell 3 of the 6dF.\\
\indent  We use the normalised correlation coefficient $R$, to determine the relative strength of any detected anisotropy signal with respect to source clustering in the shell:
\begin{equation}
R = \frac{\sum_{i} \left( N_{\rm{CR, i}} - N_{\rm{iso, i}} \right) \cdot \left(N_{\rm{M, i}} - N_{\rm{iso, i}}\right)}{\sqrt{\sum_{i}\left(N_{\rm{CR, i}} - N_{\rm{iso, i}}\right)^2} \cdot \sqrt{ \sum_{i}\left(N_{\rm{M, i}} - N_{\rm{iso, i}}\right)^2}}.
\label{eq:normalised_cross_corr_coeff}
\end{equation}
\indent In figure \ref{fig:normalised_corr_coeff} we plot the distribution of $R$ obtained in 10000 realisations of UHECR samples from an isotropic distribution of sources and from a distribution of sources based on the local galaxy distribution, as well as for the observed PAO UHECRs ($R_{\rm{Auger}}$, in shells 1, 2 and 3 from left to right, for the PSCz (\textit{top row}) and the 6dF (\textit{bottom row}). The plots shown are for $5.9^{\circ} \times 5.9^{\circ}$ angular bins. In shell 1 of the PSCz, which includes all sources nearer than 47 Mpc, the statistic $R$ has greater discriminatory power than shells 2 and 3 and the mean value of $R_{PSCz}$ in shell 1 is twice that of shells 2 and 3. The picture is similar for the 6dF. The mean value of $R_{6dF}$ drops from $\sim 0.3$ in shell 1, to $\sim 0.2$ in shell 2 and $\sim 0.1$ in shell 3. The reason we see this behaviour is that going from shell 1 to 3 there are progressively more galaxies in each shell,  occupying a larger volume and hence we move to larger scales where the galaxy distribution is less anisotropic and there is significantly less contrast. In shell 1 where $R$ has the greatest statistical power the PAO UHECRs are consistent with an isotropic hypothesis, whereas in shells 2 and 3 of the PSCz and shell 3 of the 6dF which have lower contrast and less discriminatory power there is weak evidence of correlation with the predicted source distribution. This result suggests that the true UHECR source population has similar clustering properties to those of the galaxy distribution in shells 2 and 3 and is less clustered than the source distribution in shell 1, which is consistent with the results in the other sections of this work. 
\subsection{Magnetic deflections}
\label{subsec:deflection}
We investigate the dependence of our results on the amplitude of the magnetic deflections suffered by UHECRs during their propagation. If UHECRs are Fe nuclei as some experimental data suggest \cite{2010APh....34..314T}, average deflections are likely to be of the order of $20^{\circ}$. The results obtained in this section are relevant to protons and Fe nuclei which are attenuated over similar distances at energy above $\sim 40$~EeV but not to intermediate mass nuclei which have a much shorter mean free path at this energy range (see for example \cite{2005A&A...443L..29A}). \\
\indent We generate mock realisations in which UHECRs suffer simulated magnetic deflections $d$ in the range $5^{\circ} - 20^{\circ}$ and determine the range of deflection angles for which a correlation with the source population can be established. We then compare these model distributions to the observed PAO UHECRs. We simulate random magnetic deflections in the arrival directions of UHECRs by allowing for a shift in their positions with respect to the position of their sources. Using a 2 dimensional Gaussian function with width $d$ centred at the position of the source we generate randomly oriented angular displacements for each UHECR in each mock realisation. \\
\indent In figure \ref{fig:deflection_limit} we show the distribution of values of $X$ obtained for mock UHECRs whose sources are nearby galaxies (as in our models in previous sections), and which have suffered magnetic deflections in the range $0^{\circ} - 20^{\circ}$. We find that even if the sources of the UHECRs are correlated with nearby galaxies (as the ones in our LSS model) increasing the deflection angle dilutes the anisotropy signal and for a dataset of 69 events the correlation with the source population is wiped out by deflections greater than $\sim 20^{\circ}$. We also show the value of $X$ obtained for the PAO observed UHECRs, $X_{\rm{Auger}}$. Although no models can be ruled out with present data, intermediate deflection models are favoured. 
\subsection{Systematic Uncertainties}
\label{subsec:systematics}
In the previous section we have looked at the uncertainties introduced by the unknown composition of UHECRs and the magnitude of magnetic deflections in some depth. In this section we focus on the sensitivity of our results to the uncertainty on the selection function of the galaxy surveys. \\
When modelling the selection function of a galaxy survey the uncertainty is largest nearby where the luminosity function is calculated from a very small number of galaxies. This uncertainly nearby has a non-negligible effect on our UHECR source distribution models where nearby galaxies have a greater weight than more distant ones due to flux suppression with distance and particle energy losses during propagation. In what follows we concentrate on the PSCz survey but we expect qualitatively similar findings with the 6dF survey. 
The PSCz selection function is well fit by the expression \cite{pscz}:
\begin{equation}
\psi(r) = \psi_{\star} \left(\frac{r}{r_{\star}}\right)^{(1-\alpha)} \left[ 1 + \left(\frac{r}{r_{\star}}\right)^{\gamma} \right] ^{-\left(\frac{\beta}{\gamma}\right)}
\label{eq:pscz_sel_fn}
\end{equation}
\noindent where $\psi_{\star}$, $\alpha$, $r_{\star}$, $\gamma$ and $\beta$ describe the normalisation, the nearby slope, the break distance in comoving Mpc, its sharpness and the additional slope beyond it respectively. The published values of these parameters are $\psi_{\star} = 0.0077$, $\alpha = 1.82$, $r_{\star} = 86.4$~$h^{-1}$~Mpc, $\gamma = 1.56$, $\beta = 4.43$. In table \ref{tab:systematics} we show the sensitivity of $P_{iso}$, $P_{PSCz}$ and $\frac{P(\rm{PSCz}|\rm{PAO})}{P(\rm{ISO}|\rm{PAO})}$ on the error in $\psi(r)$ by varying the parameter $\alpha$, which controls the nearby slope, so as to reproduce the published errors. Further, in the same table we show the sensitivity of these quantities on the choice of $r_{\rm{min}}$, the distance at which the selection function is normalised, the choice of which affects the predicted galaxy number density as a function of distance nearby. \\
We see that the values of $P_{\rm{iso}}$, $P_{\rm{PSCz}}$ and $\frac{P(\rm{PSCz}|\rm{PAO})}{P(\rm{ISO}|\rm{PAO})}$ obtained for shells 1 and 2 are particularly sensitive to our systematics whereas those obtained for shell 3 are very robust in comparison. This is as expected since as we've already mentioned the systematic uncertainties in the selection function are large nearby. The sensitivity of $P_{\rm{iso}}$, $P_{\rm{PSCz}}$ and $\frac{P(\rm{PSCz}|\rm{PAO})}{P(\rm{ISO}|\rm{PAO})}$ to the systematics discussed here is comparable to the sensitivity to the choice of bin size everywhere but in shell 3 where the systematic uncertainty is small. With a larger UHECR dataset the sensitivity of our method to bin size will significantly decrease but the systematic uncertainty discussed here will not, at least not for the galaxy surveys discussed. \\
\indent There is a $23\%$ systematic uncertainty in the energy determination of the primary particles detected at PAO which introduces an error to our results. We have investigated how this uncertainty affects our results, by changing $E_f$ in our models by $23\%$ i.e. assuming that the energy of the observed PAO UHECR events has been under(over)-estimated by $23\%$. This produces changes in $P_{\rm{iso}}$($P_{\rm{PSCz}}$) of order few(10)$\%$, but doesn't strongly change our conclusions. The sensitivity to the injection spectrum index \eqref{eq:injection_spectrum} is smaller, at most few $\%$ relative to the value of $P_{\rm{iso}}$($P_{\rm{PSCz}}$) in the range $-2.5 \leq \alpha \leq -1.5$.
\begin{table}
\begin{flushleft}
  \begin{tabular}{  | l | c | c | c | c | c | c | c | c | c | c | c | c |}
  \hline
   &\multicolumn{3}{c |}{Entire PSCz} &\multicolumn{3}{c |}{Shell 1} & \multicolumn{3}{c |}{Shell 2} & \multicolumn{3}{c |}{Shell 3} \\ 
       & \multirow{4}{*}{\rotatebox{90}{$P_{\rm{iso}} (\%)$}} & \multirow{4}{*}{\rotatebox{90}{$P_{\rm{PSCz}} (\%)$}} &\multirow{4}{*}{\rotatebox{90}{$\frac{P(\rm{LSS}|\rm{PAO})}{P(\rm{ISO}|\rm{PAO})}$}} & \multirow{4}{*}{\rotatebox{90}{$P_{\rm{iso}} (\%)$}} &  \multirow{4}{*}{\rotatebox{90}{$P_{\rm{PSCz}} (\%)$}} & \multirow{4}{*}{\rotatebox{90}{$\frac{P(\rm{LSS}|\rm{PAO})}{P(\rm{ISO}|\rm{PAO})}$}} &  \multirow{4}{*}{\rotatebox{90}{$P_{\rm{iso}} (\%)$}} &   \multirow{4}{*}{\rotatebox{90}{$P_{\rm{PSCz}} (\%)$}} &  \multirow{4}{*}{\rotatebox{90}{$\frac{P(\rm{LSS}|\rm{PAO})}{P(\rm{ISO}|\rm{PAO})}$}} &   \multirow{4}{*}{\rotatebox{90}{$P_{\rm{iso}} (\%)$}} &  \multirow{4}{*}{\rotatebox{90}{$P_{\rm{PSCz}} (\%)$}} & \multirow{4}{*}{\rotatebox{90}{$\frac{P(\rm{LSS}|\rm{PAO})}{P(\rm{ISO}|\rm{PAO})}$}} \\ 
     ($\alpha$,  &&&&&&&&&&&&\\
      $r_{\rm{min}}$~[Mpc])&&&&&&&&&&&&\\
     &&&&&&&&&&&&\\
     \hline 
     $1.82, 0.84$ & 6.12 & 0.61 & 0.5 & 35.1 & 0.27 & 0.1 & 2.13 & 40.9 & 7 & 1.40 & 31.5 & 10 \\ \hline
    $1.68, 0.84$  & 12.2 & 0.04 & 0.1 & 38.2 & 0.05 & 0.05 & 3.46 & 22.9 & 5 & 2.12 & 28.7 & 7 \\ \hline
    $1.94, 0.84$ & 4.09 & 2.86 &1 & 13.1 & 4.31 & 0.5 & 20.7 & 12.7 & 0.7 & 2.26 & 43.9 & 8 \\ \hline
    $1.82, 5$ & 20.2 & 0.03 & 0.06 & 48 & 0 & 0.03 & 2.46 & 40.4 & 8 & 2.54 & 29.6 & 6 \\ \hline
    $1.82, 10$ & 23.9 & 0 & 0.01 & 44.7 & 0 & 0.01 & 2.76 & 20.5 & 5 & 2.72 & 25.8 & 5 \\ \hline
\end{tabular}	
     \caption{Sensitivity of $P_{\rm{iso}}$ ($P_{\rm{PSCz}}$), the percentage of realisations of an isotropic (correlated with LSS) UHECR source distribution in which the value of $X$ was more extreme than $X_{\rm{Auger}}$, the value of $X$ obtained with the observed PAO UHECRs, and $\frac{P(\rm{PSCz}|\rm{PAO})}{P(\rm{ISO}|\rm{PAO})}$, the ratio of the likelihoods of the two models of source distribution, to errors in the PSCz selection function and to the choice of the selection function normalisation distance $r_{\rm{min}}$. From left to right we show the sensitivity for the source model based on the entire PSCZ catalogue (introduced in $\S$ \ref{subsec:cell_size}) and for shells 1, 2 and 3 as defined in $\S$ \ref{subsec:shells}. All values quoted in this table are for an analysis performed with equal area angular bins with size $5.9^{\circ} \times 5.9^{\circ}$. The top row gives the canonical choice of parameters for this analysis, the second and third row give sensitivity to the parameter $\alpha$ of \eqref{eq:pscz_sel_fn} and the two bottom rows give the sensitivity to $r_{\rm{min}}$ (introduced in  $\S$\ref{subsec:gzk_weight}). }
     \label{tab:systematics}
  \end{flushleft}
  \end{table}
\section{Discussion and Conclusions}
\label{sec:Discussion}
We have used the set of 69 observed PAO UHECRs with energy greater than $55 $~EeV to assess whether their arrival directions are correlated with the positions of nearby galaxies or sources correlated with those galaxies. We have modelled a steady UHECR source distribution in which individual UHECR sources are faint using the PSCz and 6dF which is being used here for the first time to model the predicted UHECR source distribution. A source distribution in which individual UHECR sources are faint is well motivated in the absence of a significant number of ``repeaters'', which we don't see in the 69 PAO UHECRs (see $\S$\ref{subsec:gzk_weight}). Throughout this work we have taken into account the UHECR flux suppression due to energy losses during particle propagation, expected random magnetic deflections of a few degrees and the non-uniform PAO exposure. \\
\indent In figures \ref{fig:histograms_cell_size_69_pscz}-\ref{fig:6df_pscz}, we have shown that the observed PAO UHECRs have a higher degree of correlation with the predicted UHECR source distribution than 94\% (98\%) of mock realisations from an isotropic source distribution, when cross-correlated with the PSCz (6dF), sensitive to the choice of cell-size for the analysis and other systematics discussed in $\S$\ref{subsec:systematics}. At the same time the observed cross-correlation signal is lower than in $\gtrsim 85\%$ of realisations of UHECRs that originate in galaxies in either survey. As shown in figure \ref{fig:deflection_limit}, the observed PAO UHECRs favour a model in which the sources are galaxies in the PSCz/6dF, but random magnetic deflections of UHECRs are slightly greater than in our default model parameters, of order $5-10^{\circ}$. Our results are very sensitive to the choice of cell-size due to low statistics, but this sensitivity will certainly decrease when the number of observed UHECRs increases.\\
\indent We have proposed a new way of analysing any observed correlation between observed UHECRs and model source distribution, by dividing the predicted source distribution into  radial shells of equal predicted UHECR flux contribution. We have shown that the 69 observed PAO UHECRs are consistent with an isotropic distribution when cross-correlated with the source distribution in the nearest shell (shell 1) of the PSCz and shells 1 and 2 of the 6dF, whereas there is weak evidence of correlation with the source distribution of shells 2 and 3 of the PSCz and shell 3 of the 6dF (figures \ref{fig:shells_pscz} - \ref{fig:normalised_corr_coeff}). This result suggests that the true UHECR sources have a distribution which is less clustered than the galaxy distribution in shell 1, but similar to that in shell 3 (and additionally shell 2 in the case of the PSCz.) The principle of the method we have proposed here can be further developed in future and with larger datasets, to help localise the true sources of UHECRs and constrain some of their properties.\\
\indent Our analysis has been performed with two galaxy surveys which have been carried out using different selection criteria and have different median depths, hence probing different LSS. Observing a correlation with a specific astrophysical population does not constitute proof that the sources of UHECRs are members of that population, as matter in the universe is clustered and different astrophysical populations are correlated with each other. There is very good agreement between the results obtained with the two surveys despite them having different fields of view and median depths, highlighting the robustness of the method used. \\
\indent In this work we have taken a Frequentist approach to the long standing question of the origin of UHECRs, as we consider it the most straightforward approach with the smallest number of assumptions regarding the source population. Other authors have taken a Bayesian approach to the question \cite{Mortlock}, and despite using different models of source population they have obtained results that are qualitatively similar to the ones obtained in this analysis, namely that the UHECRs observed to date are neither consistent with an isotropic distribution of sources nor with a model in which all UHECRs are protons that originate in nearby galaxies. \\
\indent Observed UHECRs may have suffered magnetic deflections larger than a few degrees if at least a fraction of them are heavily charged nuclei as opposed to nucleons which we assumed throughout most of this work, or if intervening magnetic fields are stronger than some recent works suggest, in which case even proton UHECRs may have suffered deflections larger than $~ 3^{\circ}$ (see for example \cite{2004PhRvD..70d3007S}). Identifying the mass composition of the primaries and better understanding of EGMFs will help break the degeneracy between the above scenarios.\\
\section{Acknowledgements}
\label{sec:acknowledgements}
FO and ST acknowledge the support of UCL's Institute of Origins. AC acknowledges support from the Royal Society for an International Incoming Fellowship. FBA acknowledges the support of the Royal Society via a Royal Society URF award. OL acknowledges support from a Royal Society Wolfson Research Merit Award, a Leverhulme Senior Research Fellowship and an Advanced Grant from the European Research Council. 
\bibliographystyle{JHEP}
\bibliography{bibliography}

\providecommand{\href}[2]{#2}\begingroup\raggedright\begin{thebibliography}{10}

\bibitem{2000PhR...327..109B}
P.~{Bhattacharjee}, {\it {Origin and propagation of extremely high energy
  cosmic rays}},  {\em \physrep} {\bf 327} (Mar., 2000) 109--247,
  [\href{http://xxx.lanl.gov/abs/astro-ph/9811011}{{\tt astro-ph/9811011}}].

\bibitem{2000RvMP...72..689N}
M.~{Nagano} and A.~A. {Watson}, {\it {Observations and implications of the
  ultrahigh-energy cosmic rays}},  {\em Reviews of Modern Physics} {\bf 72}
  (July, 2000) 689--732.

\bibitem{1984ARA&A..22..425H}
A.~M. {Hillas}, {\it {The Origin of Ultra-High-Energy Cosmic Rays}},  {\em
  \araa} {\bf 22} (1984) 425--444.

\bibitem{1966PhRvL..16..748G}
K.~{Greisen}, {\it {End to the Cosmic-Ray Spectrum?}},  {\em Physical Review
  Letters} {\bf 16} (Apr., 1966) 748--750.

\bibitem{Zatsepin-Kuzmin}
G.~T. {Zatsepin} and V.~A. {Kuz'min}, {\it {Upper Limit of the Spectrum of
  Cosmic Rays}},  {\em Soviet Journal of Experimental and Theoretical Physics
  Letters} {\bf 4} (Aug., 1966) 78.

\bibitem{2008PhRvL.100j1101A}
{The High Resolution Fly's Eye Collaboration}, {\it {First Observation of the
  Greisen-Zatsepin-Kuzmin Suppression}},  {\em Physical Review Letters} {\bf
  100} (Mar., 2008) 101101,
  [\href{http://xxx.lanl.gov/abs/astro-ph/0703099}{{\tt astro-ph/0703099}}].

\bibitem{2010PhLB..685..239A}
{The Pierre Auger Collaboration}, {\it {Measurement of the energy spectrum of
  cosmic rays above 10$^{18}$ eV using the Pierre Auger Observatory}},  {\em
  Physics Letters B} {\bf 685} (Mar., 2010) 239--246,
  [\href{http://xxx.lanl.gov/abs/1002.1975}{{\tt arXiv:1002.1975}}].

\bibitem{2012arXiv1205.5067A}
{The Telescope Array Collaboration}, {\it {The Cosmic Ray Energy Spectrum
  Observed with the Surface Detector of the Telescope Array Experiment}},  {\em
  ArXiv e-prints} (May, 2012) [\href{http://xxx.lanl.gov/abs/1205.5067}{{\tt
  arXiv:1205.5067}}].

\bibitem{1997ApJ...483....1W}
E.~{Waxman}, K.~B. {Fisher}, and T.~{Piran}, {\it {The Signature of a
  Correlation between Cosmic-Ray Sources above 10 19 eV and Large-Scale
  Structure}},  {\em \apj} {\bf 483} (July, 1997) 1,
  [\href{http://xxx.lanl.gov/abs/astro-ph/9604005}{{\tt astro-ph/9604005}}].

\bibitem{2010PhyU...53..691P}
K.~V. {Ptitsyna} and S.~V. {Troitsky}, {\it {Physical conditions in potential
  accelerators of ultra-high-energy cosmic rays: updated Hillas plot and
  radiation-loss constraints}},  {\em Physics Uspekhi} {\bf 53} (Oct., 2010)
  691--701, [\href{http://xxx.lanl.gov/abs/0808.0367}{{\tt arXiv:0808.0367}}].

\bibitem{2012Natur.484..351A}
{ICECUBE Collaboration}, {\it {An absence of neutrinos associated with
  cosmic-ray acceleration in {$\gamma$}-ray bursts}},  {\em \nat} {\bf 484}
  (Apr., 2012) 351--354, [\href{http://xxx.lanl.gov/abs/1204.4219}{{\tt
  arXiv:1204.4219}}].

\bibitem{2012arXiv1205.3479D}
A.~{Dar}, {\it {Neutrinos And Cosmic Rays From Gamma Ray Bursts}},  {\em ArXiv
  e-prints} (May, 2012) [\href{http://xxx.lanl.gov/abs/1205.3479}{{\tt
  arXiv:1205.3479}}].

\bibitem{2012ApJ...752...29H}
H.-N. {He}, R.-Y. {Liu}, X.-Y. {Wang}, S.~{Nagataki}, K.~{Murase}, and Z.-G.
  {Dai}, {\it {Icecube Nondetection of Gamma-Ray Bursts: Constraints on the
  Fireball Properties}},  {\em \apj} {\bf 752} (June, 2012) 29,
  [\href{http://xxx.lanl.gov/abs/1204.0857}{{\tt arXiv:1204.0857}}].

\bibitem{2010PhRvL.104i1101A}
{The Pierre Auger Collaboration}, {\it {Measurement of the Depth of Maximum of
  Extensive Air Showers above $10^{18}$~eV}},  {\em Physical Review Letters}
  {\bf 104} (Mar., 2010) 091101, [\href{http://xxx.lanl.gov/abs/1002.0699}{{\tt
  arXiv:1002.0699}}].

\bibitem{2010arXiv1010.2690S}
P.~{Sokolsky} and {for the HiRes Collaboration}, {\it {Final Results from the
  High Resolution Fly's Eye (HiRes) Experiment}},  {\em ArXiv e-prints} (Oct.,
  2010) [\href{http://xxx.lanl.gov/abs/1010.2690}{{\tt arXiv:1010.2690}}].

\bibitem{Jui:2011vm}
{The Telescope Array Collaboration}, {\it {Cosmic Ray in the Northern
  Hemisphere: Results from the Telescope Array Experiment}},  {\em ArXiv
  e-prints} (2011) [\href{http://xxx.lanl.gov/abs/1110.0133}{{\tt
  arXiv:1110.0133}}].

\bibitem{2011arXiv1101.1155W}
E.~{Waxman}, {\it {High energy cosmic ray and neutrino astronomy}},  {\em ArXiv
  e-prints} (Jan., 2011) [\href{http://xxx.lanl.gov/abs/1101.1155}{{\tt
  arXiv:1101.1155}}].

\bibitem{2010APh....34..314T}
{The Pierre Auger Collaboration}, {\it {Update on the correlation of the
  highest energy cosmic rays with nearby extragalactic matter}},  {\em
  Astroparticle Physics} {\bf 34} (Dec., 2010) 314--326,
  [\href{http://xxx.lanl.gov/abs/1009.1855}{{\tt arXiv:1009.1855}}].

\bibitem{kashti2008}
T.~{Kashti} and E.~{Waxman}, {\it {Searching for a correlation between
  cosmic-ray sources above $10^{19}$ eV and large scale structure}},  {\em
  Journal of Cosmology and Astro-Particle Physics} {\bf 5} (May, 2008) 006,
  [\href{http://xxx.lanl.gov/abs/0801.4516}{{\tt arXiv:0801.4516}}].

\bibitem{2010ApJ...716..914B}
A.~A. {Berlind}, G.~R. {Farrar}, and I.~{Zaw}, {\it {Correlations Between
  Ultrahigh Energy Cosmic Rays and Infrared-luminous Galaxies}},  {\em \apj}
  {\bf 716} (June, 2010) 914--917,
  [\href{http://xxx.lanl.gov/abs/0904.4276}{{\tt arXiv:0904.4276}}].

\bibitem{2009JCAP...06..031T}
H.~{Takami}, T.~{Nishimichi}, K.~{Yahata}, and K.~{Sato}, {\it
  {Cross-correlation between UHECR arrival distribution and large-scale
  structure}},  {\em \jcap} {\bf 6} (June, 2009) 31,
  [\href{http://xxx.lanl.gov/abs/0812.0424}{{\tt arXiv:0812.0424}}].

\bibitem{2006JCAP...01..009C}
A.~{Cuoco}, R.~{D'Abrusco}, G.~{Longo}, G.~{Miele}, and P.~D. {Serpico}, {\it
  {The footprint of large scale cosmic structure on the ultrahigh energy cosmic
  ray distribution}},  {\em Journal of Cosmology and Astro-Particle Physics}
  {\bf 1} (Jan., 2006) 9, [\href{http://xxx.lanl.gov/abs/astro-ph/0510765}{{\tt
  astro-ph/0510765}}].

\bibitem{2009JCAP...04..003K}
H.~B.~J. {Koers} and P.~{Tinyakov}, {\it {Testing large-scale (an)isotropy of
  ultra-high energy cosmic rays}},  {\em \jcap} {\bf 4} (Apr., 2009) 3,
  [\href{http://xxx.lanl.gov/abs/0812.0860}{{\tt arXiv:0812.0860}}].

\bibitem{2010ApJ...713L..64A}
{The High Resolution Fly's Eye Collaboration}, {\it {Analysis of Large-scale
  Anisotropy of Ultra-high Energy Cosmic Rays in HiRes Data}},  {\em \apjl}
  {\bf 713} (Apr., 2010) L64--L68,
  [\href{http://xxx.lanl.gov/abs/1002.1444}{{\tt arXiv:1002.1444}}].

\bibitem{2009arXiv0906.2347T}
{The Pierre Auger Collaboration}, {\it {Astrophysical Sources of Cosmic Rays
  and Related Measurements with the Pierre Auger Observatory}},  {\em ArXiv
  e-prints} (June, 2009) [\href{http://xxx.lanl.gov/abs/0906.2347}{{\tt
  arXiv:0906.2347}}].

\bibitem{2007Sci...318..938T}
{The Pierre Auger Collaboration}, {\it {Correlation of the Highest-Energy
  Cosmic Rays with Nearby Extragalactic Objects}},  {\em Science} {\bf 318}
  (Nov., 2007) 938--, [\href{http://xxx.lanl.gov/abs/0711.2256}{{\tt
  arXiv:0711.2256}}].

\bibitem{2008APh....29..188P}
{The Pierre Auger Collaboration}, {\it {Correlation of the highest-energy
  cosmic rays with the positions of nearby active galactic nuclei}},  {\em
  Astroparticle Physics} {\bf 29} (Apr., 2008) 188--204,
  [\href{http://xxx.lanl.gov/abs/0712.2843}{{\tt arXiv:0712.2843}}].

\bibitem{george_et_al}
M.~R. {George}, A.~C. {Fabian}, W.~H. {Baumgartner}, R.~F. {Mushotzky}, and
  J.~{Tueller}, {\it {On active galactic nuclei as sources of ultra-high energy
  cosmic rays}},  {\em \mnras} {\bf 388} (July, 2008) L59--L63,
  [\href{http://xxx.lanl.gov/abs/0805.2053}{{\tt arXiv:0805.2053}}].

\bibitem{2009PhRvD..80l3018P}
A.~{Pe'Er}, K.~{Murase}, and P.~{M{\'e}sz{\'a}ros}, {\it {Radio-quiet active
  galactic nuclei as possible sources of ultrahigh-energy cosmic rays}},  {\em
  \prd} {\bf 80} (Dec., 2009) 123018,
  [\href{http://xxx.lanl.gov/abs/0911.1776}{{\tt arXiv:0911.1776}}].

\bibitem{Mortlock}
L.~J. {Watson}, D.~J. {Mortlock}, and A.~H. {Jaffe}, {\it {A Bayesian analysis
  of the 27 highest energy cosmic rays detected by the Pierre Auger
  Observatory}},  {\em \mnras} {\bf 418} (Nov., 2011) 206--213,
  [\href{http://xxx.lanl.gov/abs/1010.0911}{{\tt arXiv:1010.0911}}].

\bibitem{2001JETPL..74..445T}
P.~G. {Tinyakov} and I.~I. {Tkachev}, {\it {BL Lacertae are Probable Sources of
  the Observed Ultrahigh Energy Cosmic Rays}},  {\em Soviet Journal of
  Experimental and Theoretical Physics Letters} {\bf 74} (Nov., 2001) 445--448,
  [\href{http://xxx.lanl.gov/abs/astro-ph/0102476}{{\tt astro-ph/0102476}}].

\bibitem{2010ApJS..186..378T}
J.~{Tueller}, W.~H. {Baumgartner}, C.~B. {Markwardt}, G.~K. {Skinner}, R.~F.
  {Mushotzky}, M.~{Ajello}, S.~{Barthelmy}, A.~{Beardmore}, W.~N. {Brandt},
  D.~{Burrows}, G.~{Chincarini}, S.~{Campana}, J.~{Cummings}, G.~{Cusumano},
  P.~{Evans}, E.~{Fenimore}, N.~{Gehrels}, O.~{Godet}, D.~{Grupe},
  S.~{Holland}, J.~{Kennea}, H.~A. {Krimm}, M.~{Koss}, A.~{Moretti},
  K.~{Mukai}, J.~P. {Osborne}, T.~{Okajima}, C.~{Pagani}, K.~{Page},
  D.~{Palmer}, A.~{Parsons}, D.~P. {Schneider}, T.~{Sakamoto}, R.~{Sambruna},
  G.~{Sato}, M.~{Stamatikos}, M.~{Stroh}, T.~{Ukwata}, and L.~{Winter}, {\it
  {The 22 Month Swift-BAT All-Sky Hard X-ray Survey}},  {\em \apjs} {\bf 186}
  (Feb., 2010) 378--405, [\href{http://xxx.lanl.gov/abs/0903.3037}{{\tt
  arXiv:0903.3037}}].

\bibitem{2004NIMPA.523...50A}
{The Pierre Auger Collaboration}, {\it {Properties and performance of the
  prototype instrument for the Pierre Auger Observatory}},  {\em Nuclear
  Instruments and Methods in Physics Research A} {\bf 523} (May, 2004) 50--95.

\bibitem{sommers2001}
P.~{Sommers}, {\it {Cosmic ray anisotropy analysis with a full-sky
  observatory}},  {\em Astroparticle Physics} {\bf 14} (Jan., 2001) 271--286,
  [\href{http://xxx.lanl.gov/abs/astro-ph/0004016}{{\tt astro-ph/0004016}}].

\bibitem{2011arXiv1107.4809T}
{The Pierre Auger Collaboration}, {\it {The Pierre Auger Observatory I: The
  Cosmic Ray Energy Spectrum and Related Measurements}},  {\em ArXiv e-prints}
  (July, 2011) [\href{http://xxx.lanl.gov/abs/1107.4809}{{\tt
  arXiv:1107.4809}}].

\bibitem{the6df}
D.~H. {Jones}, M.~A. {Read}, W.~{Saunders}, M.~{Colless}, T.~{Jarrett}, Q.~A.
  {Parker}, A.~P. {Fairall}, T.~{Mauch}, E.~M. {Sadler}, F.~G. {Watson},
  D.~{Burton}, L.~A. {Campbell}, P.~{Cass}, S.~M. {Croom}, J.~{Dawe},
  K.~{Fiegert}, L.~{Frankcombe}, M.~{Hartley}, J.~{Huchra}, D.~{James},
  E.~{Kirby}, O.~{Lahav}, J.~{Lucey}, G.~A. {Mamon}, L.~{Moore}, B.~A.
  {Peterson}, S.~{Prior}, D.~{Proust}, K.~{Russell}, V.~{Safouris},
  K.~{Wakamatsu}, E.~{Westra}, and M.~{Williams}, {\it {The 6dF Galaxy Survey:
  final redshift release (DR3) and southern large-scale structures}},  {\em
  MNRAS} {\bf 399} (Oct., 2009) 683--698,
  [\href{http://xxx.lanl.gov/abs/0903.5451}{{\tt arXiv:0903.5451}}].

\bibitem{pscz}
W.~{Saunders}, W.~J. {Sutherland}, S.~J. {Maddox}, O.~{Keeble}, S.~J. {Oliver},
  M.~{Rowan-Robinson}, R.~G. {McMahon}, G.~P. {Efstathiou}, H.~{Tadros},
  S.~D.~M. {White}, C.~S. {Frenk}, A.~{Carrami{\~n}ana}, and M.~R.~S.
  {Hawkins}, {\it {The PSCz catalogue}},  {\em MNRAS} {\bf 317} (Sept., 2000)
  55--63, [\href{http://xxx.lanl.gov/abs/astro-ph/0001117}{{\tt
  astro-ph/0001117}}].

\bibitem{1978A&A....65..415C}
G.~{Cavallo}, {\it {On the sources of ultra-high energy cosmic rays}},  {\em
  \aap} {\bf 65} (May, 1978) 415--419.

\bibitem{1996APh.....5..279R}
G.~E. {Romero}, J.~A. {Combi}, S.~E. {Perez Bergliaffa}, and L.~A.
  {Anchordoqui}, {\it {Centaurus A as a source of extragalactic cosmic rays
  with arrival energies well beyond the GZK cutoff}},  {\em Astroparticle
  Physics} {\bf 5} (Oct., 1996) 279--283,
  [\href{http://xxx.lanl.gov/abs/gr-qc/9511031}{{\tt gr-qc/9511031}}].

\bibitem{2001PhRvL..87h1101A}
L.~A. {Anchordoqui}, H.~{Goldberg}, and T.~J. {Weiler}, {\it {Auger Test of the
  Cen A Model of Highest Energy Cosmic Rays}},  {\em Physical Review Letters}
  {\bf 87} (Aug., 2001) 081101,
  [\href{http://xxx.lanl.gov/abs/astro-ph/0103043}{{\tt astro-ph/0103043}}].

\bibitem{2000PhRvL..85.1154D}
S.~L. {Dubovsky}, P.~G. {Tinyakov}, and I.~I. {Tkachev}, {\it {Statistics of
  Clustering of Ultrahigh Energy Cosmic Rays and the Number of Their Sources}},
   {\em Physical Review Letters} {\bf 85} (Aug., 2000) 1154--1157,
  [\href{http://xxx.lanl.gov/abs/astro-ph/0001317}{{\tt astro-ph/0001317}}].

\bibitem{2001PhRvD..63b3002F}
Z.~{Fodor} and S.~D. {Katz}, {\it {Ultrahigh energy cosmic rays from compact
  sources}},  {\em \prd} {\bf 63} (Jan., 2001) 023002,
  [\href{http://xxx.lanl.gov/abs/hep-ph/0007158}{{\tt hep-ph/0007158}}].

\bibitem{2000ApJ...542..542B}
J.~N. {Bahcall} and E.~{Waxman}, {\it {Ultra-High-Energy Cosmic Rays May Come
  from Clustered Sources}},  {\em \apj} {\bf 542} (Oct., 2000) 542--547,
  [\href{http://xxx.lanl.gov/abs/hep-ph/9912326}{{\tt hep-ph/9912326}}].

\bibitem{2010JCAP...10..013K}
K.~{Kotera}, D.~{Allard}, and A.~V. {Olinto}, {\it {Cosmogenic neutrinos:
  parameter space and detectabilty from PeV to ZeV}},  {\em \jcap} {\bf 10}
  (Oct., 2010) 13, [\href{http://xxx.lanl.gov/abs/1009.1382}{{\tt
  arXiv:1009.1382}}].

\bibitem{1987PhR...154....1B}
R.~{Blandford} and D.~{Eichler}, {\it {Particle acceleration at astrophysical
  shocks: A theory of cosmic ray origin}},  {\em \physrep} {\bf 154} (Oct.,
  1987) 1--75.

\bibitem{2001MNRAS.328..393A}
A.~{Achterberg}, Y.~A. {Gallant}, J.~G. {Kirk}, and A.~W. {Guthmann}, {\it
  {Particle acceleration by ultrarelativistic shocks: theory and simulations}},
   {\em \mnras} {\bf 328} (Dec., 2001) 393--408,
  [\href{http://xxx.lanl.gov/abs/astro-ph/0107530}{{\tt astro-ph/0107530}}].

\bibitem{2009JCAP...03..020K}
B.~{Katz}, R.~{Budnik}, and E.~{Waxman}, {\it {The energy production rate and
  the generation spectrum of UHECRs}},  {\em \jcap} {\bf 3} (Mar., 2009) 20,
  [\href{http://xxx.lanl.gov/abs/0811.3759}{{\tt arXiv:0811.3759}}].

\bibitem{2003JCAP...11..015F}
Z.~{Fodor}, S.~D. {Katz}, A.~{Ringwald}, and H.~{Tu}, {\it {Bounds on the
  cosmogenic neutrino flux}},  {\em Journal of Cosmology and Astro-Particle
  Physics} {\bf 11} (Nov., 2003) 15,
  [\href{http://xxx.lanl.gov/abs/hep-ph/0309171}{{\tt hep-ph/0309171}}].

\bibitem{2006MNRAS.373...45E}
P.~{Erdo{\u g}du}, O.~{Lahav}, J.~P. {Huchra}, M.~{Colless}, R.~M. {Cutri},
  E.~{Falco}, T.~{George}, T.~{Jarrett}, D.~H. {Jones}, L.~M. {Macri},
  J.~{Mader}, N.~{Martimbeau}, M.~A. {Pahre}, Q.~A. {Parker}, A.~{Rassat}, and
  W.~{Saunders}, {\it {Reconstructed density and velocity fields from the 2MASS
  Redshift Survey}},  {\em \mnras} {\bf 373} (Nov., 2006) 45--64,
  [\href{http://xxx.lanl.gov/abs/astro-ph/0610005}{{\tt astro-ph/0610005}}].

\bibitem{1999AJ....118..337C}
S.~{Courteau} and S.~{van den Bergh}, {\it {The Solar Motion Relative to the
  Local Group}},  {\em \aj} {\bf 118} (July, 1999) 337--345,
  [\href{http://xxx.lanl.gov/abs/astro-ph/9903298}{{\tt astro-ph/9903298}}].

\bibitem{1996ApJ...472L..89W}
E.~{Waxman} and J.~{Miralda-Escude}, {\it {Images of Bursting Sources of
  High-Energy Cosmic Rays: Effects of Magnetic Fields}},  {\em \apjl} {\bf 472}
  (Dec., 1996) L89, [\href{http://xxx.lanl.gov/abs/astro-ph/9607059}{{\tt
  astro-ph/9607059}}].

\bibitem{1994RPPh...57..325K}
P.~P. {Kronberg}, {\it {Extragalactic magnetic fields}},  {\em Reports on
  Progress in Physics} {\bf 57} (Apr., 1994) 325--382.

\bibitem{1997FCPh...19....1V}
J.~P. {Vallee}, {\it {Observations of the Magnetic Fields Inside and Outside
  the Milky Way, Starting with Globules (\~{} 1 parsec), Filaments, Clouds,
  Superbubbles, Spiral Arms, Galaxies, Superclusters, and Ending with the
  Cosmological Universe's Background Surface (at \~{} 8 Teraparsecs)}},  {\em
  \fcp} {\bf 19} (1997) 1--89.

\bibitem{2005JCAP...01..009D}
K.~{Dolag}, D.~{Grasso}, V.~{Springel}, and I.~{Tkachev}, {\it {Constrained
  simulations of the magnetic field in the local Universe and the propagation
  of ultrahigh energy cosmic rays}},  {\em \jcap} {\bf 1} (Jan., 2005) 9,
  [\href{http://xxx.lanl.gov/abs/astro-ph/0410419}{{\tt astro-ph/0410419}}].

\bibitem{2008PhRvD..77l3003K}
K.~{Kotera} and M.~{Lemoine}, {\it {Optical depth of the Universe to ultrahigh
  energy cosmic ray scattering in the magnetized large scale structure}},  {\em
  \prd} {\bf 77} (June, 2008) 123003,
  [\href{http://xxx.lanl.gov/abs/0801.1450}{{\tt arXiv:0801.1450}}].

\bibitem{2009arXiv0907.5194A}
R.~{Aloisio}, V.~{Berezinsky}, and A.~{Gazizov}, {\it {Ultra High Energy Cosmic
  Rays: The disappointing model}},  {\em ArXiv e-prints} (July, 2009)
  [\href{http://xxx.lanl.gov/abs/0907.5194}{{\tt arXiv:0907.5194}}].

\bibitem{2009JCAP...11..009L}
M.~{Lemoine} and E.~{Waxman}, {\it {Anisotropy vs chemical composition at
  ultra-high energies}},  {\em \jcap} {\bf 11} (Nov., 2009) 9,
  [\href{http://xxx.lanl.gov/abs/0907.1354}{{\tt arXiv:0907.1354}}].

\bibitem{2011arXiv1106.3048T}
{The Pierre Auger Collaboration}, {\it {Anisotropy and chemical composition of
  ultra-high energy cosmic rays using arrival directions measured by the Pierre
  Auger Observatory}},  {\em ArXiv e-prints} (June, 2011)
  [\href{http://xxx.lanl.gov/abs/1106.3048}{{\tt arXiv:1106.3048}}].

\bibitem{1996ApJ...468..214T}
M.~{Tegmark}, D.~H. {Hartmann}, M.~S. {Briggs}, and C.~A. {Meegan}, {\it {The
  Angular Power Spectrum of BATSE 3B Gamma-Ray Bursts}},  {\em \apj} {\bf 468}
  (Sept., 1996) 214, [\href{http://xxx.lanl.gov/abs/astro-ph/9510129}{{\tt
  astro-ph/9510129}}].

\bibitem{2006APh....26...10K}
M.~{Kachelrie{\ss}} and D.~V. {Semikoz}, {\it {Clustering of ultra-high energy
  cosmic ray arrival directions on medium scales}},  {\em Astroparticle
  Physics} {\bf 26} (Aug., 2006) 10--15,
  [\href{http://xxx.lanl.gov/abs/astro-ph/0512498}{{\tt astro-ph/0512498}}].

\bibitem{2005ApJ...622..759G}
K.~M. {G{\'o}rski}, E.~{Hivon}, A.~J. {Banday}, B.~D. {Wandelt}, F.~K.
  {Hansen}, M.~{Reinecke}, and M.~{Bartelmann}, {\it {HEALPix: A Framework for
  High-Resolution Discretization and Fast Analysis of Data Distributed on the
  Sphere}},  {\em \apj} {\bf 622} (Apr., 2005) 759--771,
  [\href{http://xxx.lanl.gov/abs/astro-ph/0409513}{{\tt astro-ph/0409513}}].

\bibitem{2005A&A...443L..29A}
D.~{Allard}, E.~{Parizot}, A.~V. {Olinto}, E.~{Khan}, and S.~{Goriely}, {\it
  {UHE nuclei propagation and the interpretation of the ankle in the cosmic-ray
  spectrum}},  {\em \aap} {\bf 443} (Dec., 2005) L29--L32,
  [\href{http://xxx.lanl.gov/abs/astro-ph/0505566}{{\tt astro-ph/0505566}}].

\bibitem{2004PhRvD..70d3007S}
G.~{Sigl}, F.~{Miniati}, and T.~A. {En{\ss}lin}, {\it {Ultrahigh energy cosmic
  ray probes of large scale structure and magnetic fields}},  {\em \prd} {\bf
  70} (Aug., 2004) 043007, [\href{http://xxx.lanl.gov/abs/0401084}{{\tt
  0401084}}].

\end{thebibliography}\endgroup
\end{document}